\documentclass[acmsmall, nonacm]{acmart}

\setcopyright{cc}
\setcctype{by}
\acmJournal{PACMHCI}
\acmYear{2026} \acmVolume{10} \acmNumber{2} \acmArticle{CSCW031}
\acmMonth{4} \acmPrice{} \acmDOI{10.1145/3788067}

\usepackage{color, colortbl}
\usepackage{subcaption}
\usepackage{longtable}
\usepackage{xcolor}
\usepackage{ulem}

\begin{document}

\title[Available, Compact, and Aligned]{People Can Accurately Predict Behavior of Complex Algorithms That Are Available, Compact, and Aligned}

\author{Lindsay Popowski}
\email{popowski@cs.stanford.edu}
\affiliation{%
  \institution{Stanford University}
  \city{Stanford}
  \state{California}
  \country{USA}
}

\author{Helena Vasconcelos}
\email{helenav@cs.stanford.edu}
\affiliation{%
  \institution{Stanford University}
  \city{Stanford}
  \state{California}
  \country{USA}
}

\author{Ignacio Javier Fernandez}
\affiliation{%
  \institution{Stanford University}
  \city{Stanford}
  \state{California}
  \country{USA}
}

\author{Chijioke Chinaza Mgbahurike}
\affiliation{%
  \institution{Stanford University}
  \city{Stanford}
  \state{California}
  \country{USA}
}

\author{Ralf Herbrich}
\email{ralf.herbrich@hpi.de}
\affiliation{%
  \institution{Hasso Plattner Institut}
  \city{Potsdam}
  \state{Brandenburg}
  \country{Germany}
  }

\author{Jeffrey Hancock}
\email{hancockj@stanford.edu}
\affiliation{%
  \institution{Stanford University}
  \city{Stanford}
  \state{California}
  \country{USA}
}

\author{Michael S. Bernstein}
\email{msb@cs.stanford.edu}
\affiliation{%
  \institution{Stanford University}
  \city{Stanford}
  \state{California}
  \country{USA}
}

\renewcommand{\shortauthors}{Popowski, et al.}

\begin{abstract}
  Users trust algorithms more when they can predict the algorithms' behavior. Simple algorithms trivially yield predictively accurate mental models, but modern AI algorithms have often been assumed too complex for people to build predictive mental models, especially in the social media domain. In this paper, we describe conditions under which even complex algorithms can yield predictive mental models, opening up opportunities for a broader set of human-centered algorithms. We theorize that users will form an accurate predictive mental model of an algorithm's behavior if and only if the algorithm simultaneously satisfies three criteria: (1)~cognitive availability of the underlying concepts being modeled, (2)~concept compactness (does it form a single cognitive construct?), and (3)~high alignment between the person's and algorithm's execution of the concept. We evaluate this theory through a pre-registered experiment ($N=1250$) where users predict behavior of 25 social media feed ranking algorithms that vary on these criteria. We find that even complex (e.g., LLM-based) algorithms enjoy accurate prediction rates when they meet all criteria, and even simple (e.g., basic term count) algorithms fail to be predictable when a single criterion fails. We also find that these criteria determine outcomes beyond prediction accuracy, such as which mental models users deploy to make their predictions.

\end{abstract}



\begin{teaserfigure}
  \centering
    \includegraphics[width=\textwidth]{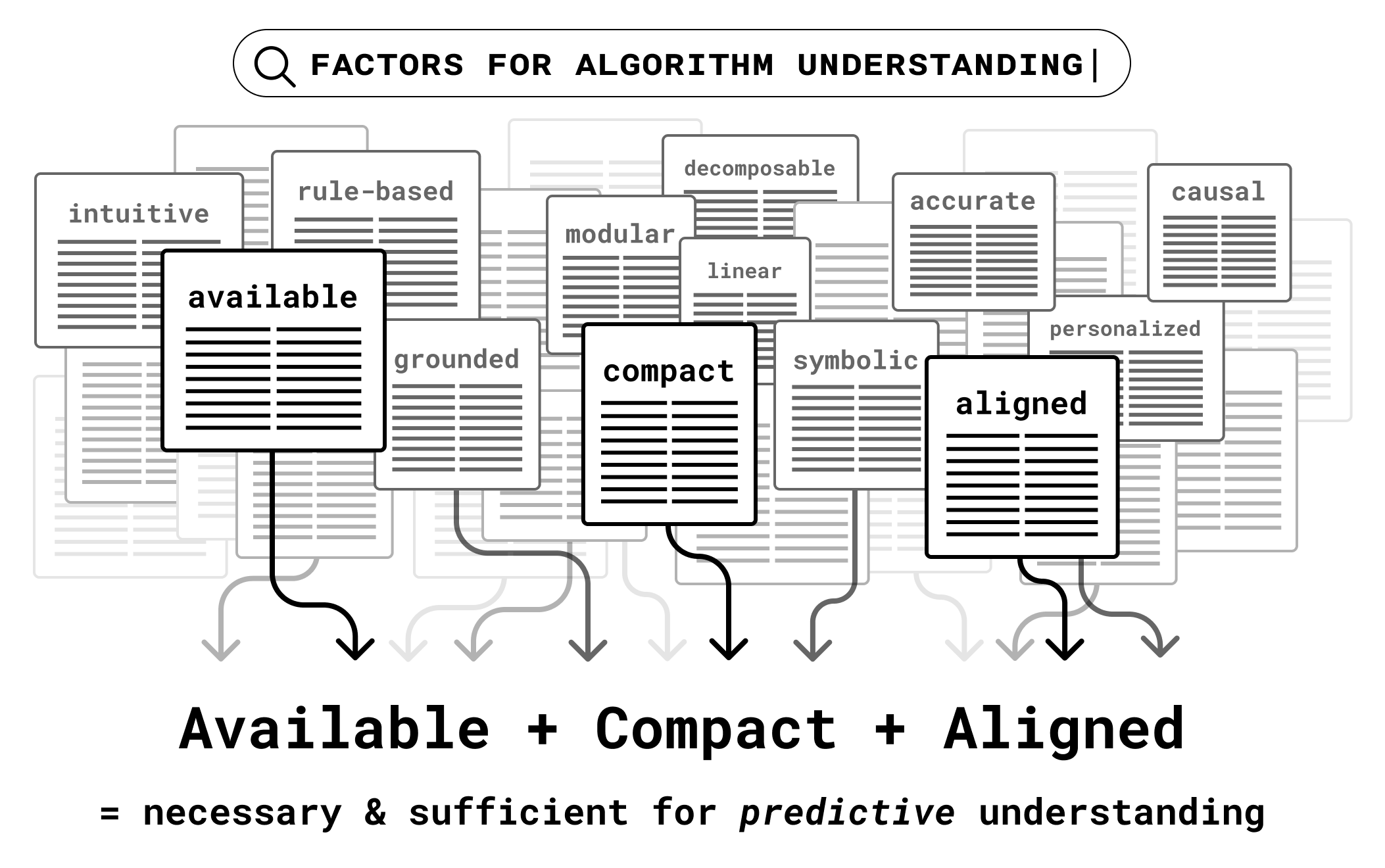}
    \Description[A search for factors for algorithm understanding results in multiple terms, including available, compact, and aligned. These are shown to be necessary and sufficient.]{A search for factors for algorithm understanding results in multiple terms displayed as documents, including available, compact, and aligned. These are shown to be necessary and sufficient. Other, similar terms are shown in the background faded, like intuitive, rule-based, grounded, modular, linear, decomposable, accurate, symbolic, causal, and personalized.}
  \caption{We propose that for an algorithm to be predictable, it must be available, compact, and aligned; and that these characteristics form a complete set that determines its predictability.}
  \label{fig: main theory performance fig}
\end{teaserfigure}


\received{May 2025}
\received[revised]{November 2025}
\received[accepted]{December 2025}

\maketitle

\section{Introduction}
Users' ability to understand---and especially, predict---the behavior of algorithms is integral to user trust in algorithmic systems~\cite{weld2019challenge}. Algorithms will therefore struggle to gain willing adoption so long as their behavior is difficult to predict. These algorithms are proposed for usage in contexts ranging from decision making (e.g.,~\cite{buccinca2020proxy, vasconcelos2023xai, barredo2020xai, weld2019challenge}) to hiring~\cite{raghavan2020hiring}, bail and parole~\cite{dressel2018accuracy}, rent-setting~\cite{vogell2022rent}, school matching~\cite{robertson2023preference}, information seeking~\cite{brin1998anatomy}, commerce~\cite{linden2003amazon}, online dating~\cite{sharabi2022finding}, job-seeking~\cite{imana2021auditing}, and displaying political content online~\cite{narayanan2023}. Here, users and the broader public have a vested interest in being able to predict how algorithms will behave, so they know how they will be impacted and can make effective choices about how and whether to interact with the algorithmic systems.

It would be no overstatement to claim that users have voiced feelings of diminished agency in interacting with these complex algorithms~\cite{eslami2016like, lee2024mindsets, alizadeh2024matchmaker} and a desire for more simple and transparent algorithms~\cite{devito2017riptwitter}. In response, artificial intelligence (AI) critics and practitioners alike have called for a turn towards interpretable and explainable AI~\cite{barredo2020xai} to provide greater agency to both algorithm users and subjects when deciding how they want to interact with algorithmic systems. 

One obvious way to ensure algorithms are predictable is to simplify them so that people can form an exact understanding of their operation and behavior. A simple algorithm, such as ranking a social media feed reverse-chronologically instead of by the usual complex set of engagement signals, is trivially easy to understand because the precise operation is so straightforward. Indeed, a significant aim within the field of interpretability is to create simpler algorithms towards this aim of better human understanding~\cite{weld2019challenge}. However, restricting ourselves to trivially simple algorithms means that we sacrifice a wide swath of higher-complexity algorithms that can achieve aims the simple ones are incapable of in order to achieve predictability and trust. 

This bias away from complex algorithms and toward simpler ones arises because people struggle to build predictive \textit{mental models} of complex algorithms~\cite{eslami2016like,french2017folk}. Mental models are internal cognitive representations of the behavior of designed systems~\cite{carroll1988mental}. They encapsulate the parts of a system that users care to understand in order to perform desired tasks~\cite{payne2007mental} such as predicting future behavior. Because a mental model is focused on what is necessary for a user to complete a task, a mental model does not need to accurately represent a system's internal operation or implementation: you don't need to know exactly how the thermodynamic refrigeration cycle works in order to build a mental model of your refrigerator's controls~\cite{norman2013design}. However, a mental model \textit{does} need to allow the user to simulate counterfactual what-ifs about how their inputs might cause the system to react. While the human-computer interaction literature is rife with examples of effective and ineffective mental models (e.g.,~\cite{norman2013design}), empirical evidence suggests that people especially struggle to build predictive mental models of complex algorithms: instead of a mental model's structured cognitive representation, when faced with complex algorithms, people often fall back on informal, incomplete beliefs called folk theories~\cite{eslami2016like,french2017folk,ytre2021folk,devito2017riptwitter}.
Taken together, these results suggest a tempting conclusion that complex algorithms are fundamentally in tension with the formation of accurate predictive mental models. 

However, complex algorithms \textit{do not necessarily} lead to error-prone mental models. This subtlety is well understood in the interpretability and explainability communities, which have pursued the question of how to help people understand complex algorithm behavior~\cite{rong2024humancenteredexplainableaisurvey, barredo2020xai}. We suggest that people can predict the behavior of even highly complex artificial intelligence algorithms, so long as those algorithms align with concepts that people can understand and replicate. 
For example, you can quite likely accurately predict the behavior of a large language model that has been asked to classify whether a particular social media post is about politics or not, despite the fact that the large language model relies on a complex attention network and hundreds of billions of parameters, while you rely on an entirely different architecture: your brain.

We synthesize prior work to articulate a theory that people create an accurate predictive mental model of an algorithm if and only if the algorithm simultaneously fulfills three criteria: the algorithm must be be \textit{Available}, \textit{Compact}, and \textit{Aligned}. These ACA criteria together capture what it means for a person to be able to build a cognitive representation of the concept underlying an algorithm's behavior: (1)~\textit{Availability}, a reference to availability bias~\cite{tversky1973availability}, captures the recognizability of the underlying concept that the algorithm is modeling. (2)~\textit{Compactness}, drawing on the literature of cognitive chunking~\cite{chase1973perception,thalmann2019does}, refers whether the algorithm's behavior can be synthesized into a single cohesive concept: is the algorithm representing a single concept or fitting together multiple concepts into a greater whole that people understand? 
(3)~Finally, \textit{Alignment} tests whether the algorithm's execution of its concept agrees with the person's execution of that concept, similar to representational alignment~\cite{sucholutsky2024representationalalignment}.
In this paper, we demonstrate empirical evidence that algorithms that fulfill all three of these criteria can produce highly accurate mental models of the algorithm's global behavior.

This theory predicts that the ACA criteria are necessary and sufficient---that no other criteria are required, but if even one fails, people will not be able to accurately predict the behavior of the algorithm. To present a test of this ACA prediction, we report an experiment ($N=1250$) where we test whether people can predict algorithm behavior across a variety of algorithms that satisfy or violate the ACA criteria. We situate this study in the domain of social media feed algorithms, which has been regularly critiqued for complex algorithms that give rise to ineffective folk theories~\cite{eslami2016like, lee2024mindsets, devito2017riptwitter}. Participants are shown a feed of political posts ranked according to a randomly assigned algorithmic condition, able to see the ranking but having no information about the algorithm underlying that ranking. They are then asked to perform a set of ranking tasks to predict which of two previously unseen posts that algorithm would rank higher. We measure whether participants correctly predict the algorithm's ranking decision. As hypothesized, we observe that participants have the highest ranking accuracy on algorithms that satisfy all three criteria, and near-random accuracy on algorithms that fail one or more criteria. Through considering the self-reported mental models of our participants, we also learn how participants gravitate towards mental models that satisfy the criteria and trade off between them when they cannot satisfy all three.

Through this Available-Compact-Aligned (ACA) theory and its associated empirical findings, we demonstrate the possibility of deploying even extremely complex algorithms that achieve high performance while remaining predictable by end-users. This design direction opens the door to a new class of algorithmic systems that can be systematically developed to be predictable to their audience, and can be tested for ``ACA compliance'' before launching. These algorithms could also assist existing efforts for participatory design for algorithms~\cite{lee2019webuild} through their human-legible representations for processes like sensemaking and deliberation \cite{lam2023sketch}. At the same time, our theory helps explain folk theorization behaviors, such as why users continue use of theories that conflict with algorithm behavior~\cite{devito2021adaptive}, maintain multiple conflicting theories~\cite{devito2018form, eslami2015close}, and adopt theories from other users~\cite{devito2021adaptive, bishop2019gossip}.

Our contributions are threefold. First, we synthesize a theory combining cognitive science with insights from interpretability, explainability, mental model formation, and folk theorization to explain the characteristics of algorithms that allow users to form accurate predictive mental models. Second, we develop a methodology for measuring the predictive accuracy of mental models for social media algorithms. Finally, using this methodology in a large-scale experimental study that includes both quantitative and qualitative analysis, we demonstrate the theory's predictive efficacy for different algorithms.

\section{Related Work}

As complex algorithms play a larger role in day-to-day life, algorithm understandability has been raised as a major concern~\cite{barredo2020xai, weld2019challenge, lipton2017mythos}.
Without users sufficiently understanding algorithms, harms emerge from errors, overreliance, and misinterpretation~\cite{karizat2021algorithmic, sandvig2016automation}. 
If we take the perspective that complex algorithms are fundamentally opaque and difficult to understand or predict, then the only way to reduce these harms is to replace complicated algorithms with simpler, more transparent ones, and accept whatever performance losses correspond to the change. 
In this paper, we work to differentiate what is computationally complex for an algorithm to execute and what is cognitively complex for a human to understand.
We aim to develop a framework for predicting the cognitive complexity of algorithm behavior, building on the extensive theory developed within the domains of explainable and interpretable artificial intelligence. 

In this section, we start by outlining the literature in AI interpretability and explainability that we draw on. We then explain our methodological bases in mental model and folk theory orientations towards user understanding of systems. We conclude by explaining our site of investigation: social media feed algorithms.

\subsection{Explainable and Interpretable AI}

Like the explainable and interpretable AI literatures, we concern ourselves with how well people can reliably understand and interpret the behavior of algorithms. Explainable and interpretable AI research is focused on creating and evaluating methods that operate over AI models to help users understand them: by offering post-hoc explanations for decisions made by algorithms (e.g., LIME~\cite{ribeiro2016lime}) and evaluating their efficacy~\cite{garreau2020explaining}; manipulating the models directly such that they operate over human understandable concepts (e.g., Concept Bottleneck Models~\cite{koh2020conceptbottleneckmodels}) and evaluating the efficacy of these inherently more interpretable models~\cite{lage2019human}; and giving global overviews of models' abilities, as in the case of a number of HCI frameworks or systems~\cite{mitchell2019cards, cabrera2023descriptions, plumbtowards, kay2016uncertainty, cabrera2023zeno}, as well as evaluating the efficacy of these systems~\cite{cabrera2023descriptions, kay2016uncertainty, cabrera2023zeno}. 
We draw heavily on this literature, particularly that regarding how to produce better human understanding of AI systems, to develop our framework, beginning with theories on what constitutes an effective explanation or interpretable model. 

 We are not the only researchers to note that simply switching to a simpler model architecture is neither necessary nor sufficient to produce user understanding. Linear models can be more opaque than even deep learning models due to the more convoluted features used in them~\cite{lipton2017mythos}. What matters is the \textit{cognitive}, rather than \textit{computational}, simplicity.
Cognitive effort similarly plays a major role in whether people verify and override AI's erroneous decisions, as well as in whether people will attempt to understand explanations~\cite{vasconcelos2023xai, buccinca2020proxy, bansal2021whole, bussone205explanation}. Even explanations that are simple compared to the original model may not be simple enough for the human to accept, so projects have aimed to lower the amount of cognitive effort required for people to understand explanations or use them to verify the model output~\cite{abdul2020cogam, ribeiro2016lime, Ustun_2015}. Our theory focuses on what characteristics of algorithms allow users to form, hold, and use mental models with minimal cognitive effort. We are influenced especially by model architectures that use concepts as a building component to render themselves more interpretable to users~\cite{lam2023sketch, koh2020conceptbottleneckmodels}. We recognize \textit{concepts} as the building block of user mental models, and therefore base our conception of cognitive simplicity in terms of concepts, rather than arbitrary features.

However, concepts are not universal: one user's understanding of a particular concept is likely to differ from another user's, while a third user could have no familiarity~\cite{chi1981representation, yee2016putting, rosch1978categorization}. Thinking about algorithms in terms of conceptual simplicity and clarity, while necessary to examine user understanding, introduces an additional layer of subjectivity. Our theory must encompass the factors that influence users' differing conceptions of the same underlying concept. In other human-AI interaction contexts, people's understanding is impacted by factors like the social context and users' intuition formed from experience~\cite{chen2023intuition, ehsan2021expanding, chen2023explanation}. We therefore incorporate psychological, experiential, and social factors into the key criteria of our theory to explain different levels of concept awareness and conceptualization across user mental models.

\subsection{Understanding User Understanding via Mental Models and Folk Theories}

In human-computer interaction (HCI), mental models are a framework for thinking of how users understand the behavior of a system ~\cite{payne2007mental, carroll1988mental}. Mental models are a structure that reflects ``the user's understanding of what the system contains, how it works, and why it works that way''~\cite{carroll1988mental}, first and foremost a framework for understanding user \textit{behavior} with regards to systems~\cite{payne2007mental}.
Mental models therefore need not describe the internals of a system, but rather should capture the overall behavior of interest. Mental models are tailored to particular tasks or needs of a user, and should be measured in their utility towards a particular task. Due to both the task specificity and the variability of users, many different mental models can exist for the same system~\cite{payne2007mental}.

 To contend with this multiplicity, HCI recommends that system designers designate an intended mental model and try to communicate it through the design or even straightforward teaching~\cite{payne2007mental}---different presentations of output can significantly affect user mental models, agnostic to the underlying models or mechanisms~\cite{wang2025chatbot}. The intended mental model should guide the user to interact with the system in the optimal manner. While mental models can describe basically any cognitive representation for a system, they are often thought of in a structured input-to-output format and measured quantitatively. 

Lay users' ability to make sense of complex algorithmic systems without fully understanding their mechanistic workings has also been studied through the lens of ``folk science,'' where the resulting folk theories are understood to be informal and oft-incomplete while still offering predictive value~\cite{keil2010folkfeasibility, rozenblit2002folklimits}. Folk theories are an alternate way of understanding users' frameworks to reason about complex systems, encompassing a broader range of possible concepts than mental models, including ideas about the motivation of the system's builders (e.g., that the platform's goals are in conflict with their own~\cite{alizadeh2024matchmaker}) and the materials or tools used in construction ~\cite{french2017folk}. Users approach their theorization in myriad ways---with differing complexity levels and structures, from basic awareness to full mechanistic descriptions; varying information sources, both endogenous and exogenous; and eventually, differing levels of success when deployed by users to interact on social media platforms~\cite{devito2021adaptive}.

Social media algorithms are the site of massive folk-theorization efforts: from developing theories how to exert control~\cite{eslami2016like}, to theorizing that they restrict certain identities~\cite{karizat2021algorithmic}, to noting how they reflect one's own personality and interests via individualization~\cite{lee2022algcrystal}, to resisting change from simple chronological feeds out of fear for negative changes~\cite{devito2017riptwitter}. Much of our knowledge of user understanding of social media algorithms comes from these depictions, which highlight user anxieties, struggles, and stresses~\cite{devito2017riptwitter, karizat2021algorithmic, eslami2016like}. 
This type of investigation yields design implications for platforms related to the interface or communication with users like incorporating seamful design~\cite{eslami2016like}, improving misleading interfaces~\cite{mayworm2024spirit}, and to educate users on how the platform works~\cite{devito2021adaptive}. 
Folk theory elicitation lets us know what users' perceptions of these algorithms might be, and produces design recommendations accordingly. 
While implications occasionally point to algorithm behavior, like limiting algorithm change~\cite{devito2021adaptive, devito2017riptwitter} or fixing algorithm behavior that creates perception of discrimination or harm~\cite{mayworm2024spirit}, this type of recommendation is not the focus. 
Folk theory elicitation is concerned more with what folk theories imply about user perceptions than whether the theories are correct~\cite{mayworm2024spirit}, and notably does not identify misperceptions of the algorithm behavior. 
So, while this method is well-equipped to highlight issues for users, it does not aim to provide the answer on how to address those issues in terms of technical algorithm operations. 
Our theory development efforts focus on this question of \textit{accuracy}, where users' understanding is measured in terms of predictive capacity towards the algorithm behavior. 
We aim to enable designers and practitioners to produce algorithmic systems that afford accurate user understanding.
At the same time, we can better understand the practices users employ when fully accurate theories are not possible, particularly in terms of the tradeoffs they have to make, speaking back to folk theories.

Our work thereby seeks to bridge the methodologies and perspectives of folk theory elicitation and mental model measurement, demonstrating that their findings complement each other. Whereas in the human-AI partnership context, people are asked to detect AI errors or make optimal decisions, prior work has shown that this task is distinct from that of AI behavior prediction~\cite{buccinca2020proxy}. Meanwhile folk theories of social media feeds are developed to model a broader set of behaviors, which include both general behavior understanding, prediction, and explaining unusual behavior, which can help explain the discrepancies in human understanding across these cases.

We combine the generally quantitative methodology of mental model measurement with the qualitative direction of folk theory elicitation to help reconcile the two directions. Rather than completely abstract away the mental representations that users are forming to focus on task performance, we collect user folk theory concepts---intended to help capture how they are mentally modeling the algorithm behavior---and analyze these representations to see if they align with the intended algorithm behavior.

\subsection{Social Media Algorithms}

Social media feed algorithms---the logic behind which posts users see on social media, and in what order---are the domain of our investigation into user understanding of algorithms. We choose this context as a promising site to understand organic mental model formation by non-technical and non-expert users. Unlike many other human-AI interaction contexts, social media users are not trained in the technology behind AI and also rarely have technical knowledge that makes them uniquely trained to understand social media (except perhaps for influencers or similar media professionals), but still care deeply about the content that social media selects for them to consume. 

On most widely-used platforms today, the feed algorithm of choice optimizes for some combination of engagement-based features, including likes, shares, comments, views, dwell-time, and other measurements that help represent user approval~\cite{ciampaglia2018algorithmic, milli2021optimizingengagement}. They also commonly incorporate some sort of time decay, ensuring that new content is prioritized over old. Additionally, researchers have proposed optimizing social media algorithms for more pro-social outcomes instead, such as democratic values~\cite{jia2024embedding}. However, many users harken back to perceived better days, when all algorithms were just a simple reverse chronological ordering~\cite{devito2017riptwitter}, mirroring the calls for more transparent and simple algorithms in other contexts. Platforms do not tend to explain the exact behavior of their algorithms, leaving users to instead theorize how the algorithm works~\cite{eslami2016like} and contributing to users' feelings of animosity towards the introduction of complex algorithms~\cite{devito2017riptwitter}. Users often find social media algorithms to be discriminatory~\cite{2025williamsblack, devito2022transfem, karizat2021algorithmic}, unpredictable~\cite{eslami2016like, devito2022transfem, devito2018form}, and unintuitive~\cite{eslami2016like, devito2021adaptive, devito2018form}, causing users to struggle to adapt their posting, sharing, and browsing habits to achieve their desired goals~\cite{devito2021adaptive, devito2018form} or resist undesired algorithmic outcomes~\cite{2025williamsblack, karizat2021algorithmic} despite extensive theorization effort.

Meanwhile, these opaque algorithms are a site of social and political significance. They are widely theorized to have played a role in political polarization~\cite{milli2023engagementusersatisfactionamplification}, and are known to impact people's beliefs~\cite{brady2023overperception}, emotions and well-being~\cite{hancock2022psychological, kreski2021social}, and behaviors~\cite{vannucci2020social}. In addition to these effects on individuals, they have the potential for magnified impacts through mass-usage and user interactions. People therefore care deeply about algorithm behavior in this context; it has the potential to impact their lives. 

Proposed improvements in this space point in different directions. Researchers have proposed optimizing social media algorithms for more pro-social outcomes instead, such as prioritizing pro-democratic content~\cite{jia2024embedding} or ranking based on different shared values~\cite{kolluri2025alexandria}. Many users instead harken back to perceived better days, when all algorithms were just a simple reverse chronological ordering~\cite{devito2017riptwitter}, mirroring the calls for more transparent and simple algorithms in other contexts. While opacity and unpredictability are not users' only concerns about modern-day feed algorithms, they are clearly one factor that matters for algorithm trust and acceptance~\cite{weld2019challenge, hong2020human, devito2021adaptive}, and particularly help users appropriately calibrate their trust~\cite{kaur2020interpreting}. Our work seeks to demonstrate the feasibility of a new class of feed algorithm, one that has similar intelligibility for users as chronological feeds while also achieving parallel aims that require more complex internal behavior: personalizing to unique user interests, filtering away harmful content, and even tackling higher-level societal ills.

\section{ACA: Available, Compact, and Aligned} \label{theory}

\begin{quote}
    \textit{``In this path toward performance, when the performance comes hand in hand with complexity, interpretability encounters itself on a downwards slope that until now appeared unavoidable.''}~\cite{barredo2020xai}
\end{quote}

It seems intuitive to assume an inherent trade-off between the complexity of an algorithm and people’s ability to predict that algorithm's behavior. In this context, we use \textit{complex} and \textit{simple} as Weberian ideal types~\cite{weber1925wirtschaft} of an algorithm to describe what others have termed comprehensibility~\cite{gleicher2016framework} of the algorithm. 
Sull and Eisenhardt distinguish simple from complex rules by being few and concise vs. many and verbose~\cite{sull2015simple}. 
In our context, machine learning algorithms are often criticized as being so complex as to be uninterpretable~\cite{weld2019challenge}, whereas a simple algorithm like selecting the most recent item in a database is easy to predict and reason about counterfactuals~\cite{weld2019challenge}. In response, much of the AI interpretability field has pursued local or global simplifications of complex models, arguing that any harms to performance are offset by the improved trust, predictability, and reliability of simpler systems~\cite{barredo2020xai}.

Our argument is that this intuition is not correct: complexity is not inherently in tension with predictive mental models. Clearly, as an algorithm gets more complex, a person's mental model can no longer match the underlying system implementation~\cite{norman2013design}: nobody memorizes hundreds of billions of parameter weights in a large language model. That does not mean, however, that a mental model cannot be predictively accurate: in our lives, we predict many of the behaviors of gravity without fully understanding general relativity. 
Likewise, a deep learning model has an extremely complex internal implementation, but if its task is to identify dogs vs. cats, we can build a mental model that accurately predicts its behavior. Likewise, we often think of a reverse chronological social media algorithm as the only one that we can understand and predict, but we could also predict the behavior of an extremely complex algorithm that sorts posts about food to the top and all others to the bottom.

In this section, we derive ACA theory---Available, Compact, and Aligned---synthesizing from foundational cognitive theory and prior work in order to predict the key conditions under which people build accurate predictive models of algorithmic behavior. We organize ACA theory in terms of the four building blocks of a theory~\cite{whetten1989theory, bhattacherjee2012theory}: constructs, propositions, logic, and boundary conditions. Constructs include the criteria of the theory, while the proposition explains the relationship between the constructs. Logic describes the reasoning behind the theory, while boundary conditions explain the underlying assumptions. 
We will revisit boundary conditions in the Discussion section so that they can accompany engagement with empirical limitations.

\subsection{Motivating Logic: How We Form Mental Models}
From the mental models literature dating back to Johnson-Laird and Norman in the 1980s~\cite{johnson1983mental, norman2014some}, there is a structured process that we use when we build mental models: (1)~recognize system affordances, (2)~represent those affordances into a mental model, and (3)~apply that mental model to predict system behavior. These three stages apply in a loop: system responses to our inputs produce new information that we recognize, update our representation, and then apply again.

If any of these three stages is absent, then the user will not be able to predict the system's behavior. If we cannot recognize the affordances, then the system's behavior remains mysterious; if we can recognize affordances but cannot build a coherent representation out of those affordances, then we fail to integrate our observations into a usable mental model; if we can recognize affordances and represent them coherently but produce incorrect predictions when applying our resulting mental model, then the entire process has lost its predictive utility. However, recognition, representation, and application are human behaviors, and do not imply anything specific about how to design \textit{algorithms} that support those behaviors. So, we must identify algorithmic criteria that ensure each of those behaviors is achievable. 

Since these three stages are a complete set, we derive one algorithmic criterion that gates the completion of each of these three stages: availability to ensure recognizability of affordances, compactness to ensure representability of those affordances into a mental model, and alignment to ensure that our application of the mental model matches the algorithm's behavior. These criterion become the constructs of our theory. By construction, then, if all three criteria are simultaneously met, people should be able to predict the algorithm's behavior---but if one of the three criteria fails, then the user will likewise fail to recognize, represent, or apply the mental model, barring them from producing a correct prediction of algorithm behavior. 

\begin{quote}
\textsc{Core ACA Proposition}: \textit{People accurately predict the behavior of an algorithm if and only if the algorithm satisfies all three criteria: that it is available, compact, and aligned.}
\end{quote}

\begin{figure}[tb]
  \centering
    \includegraphics[width=0.9\textwidth]{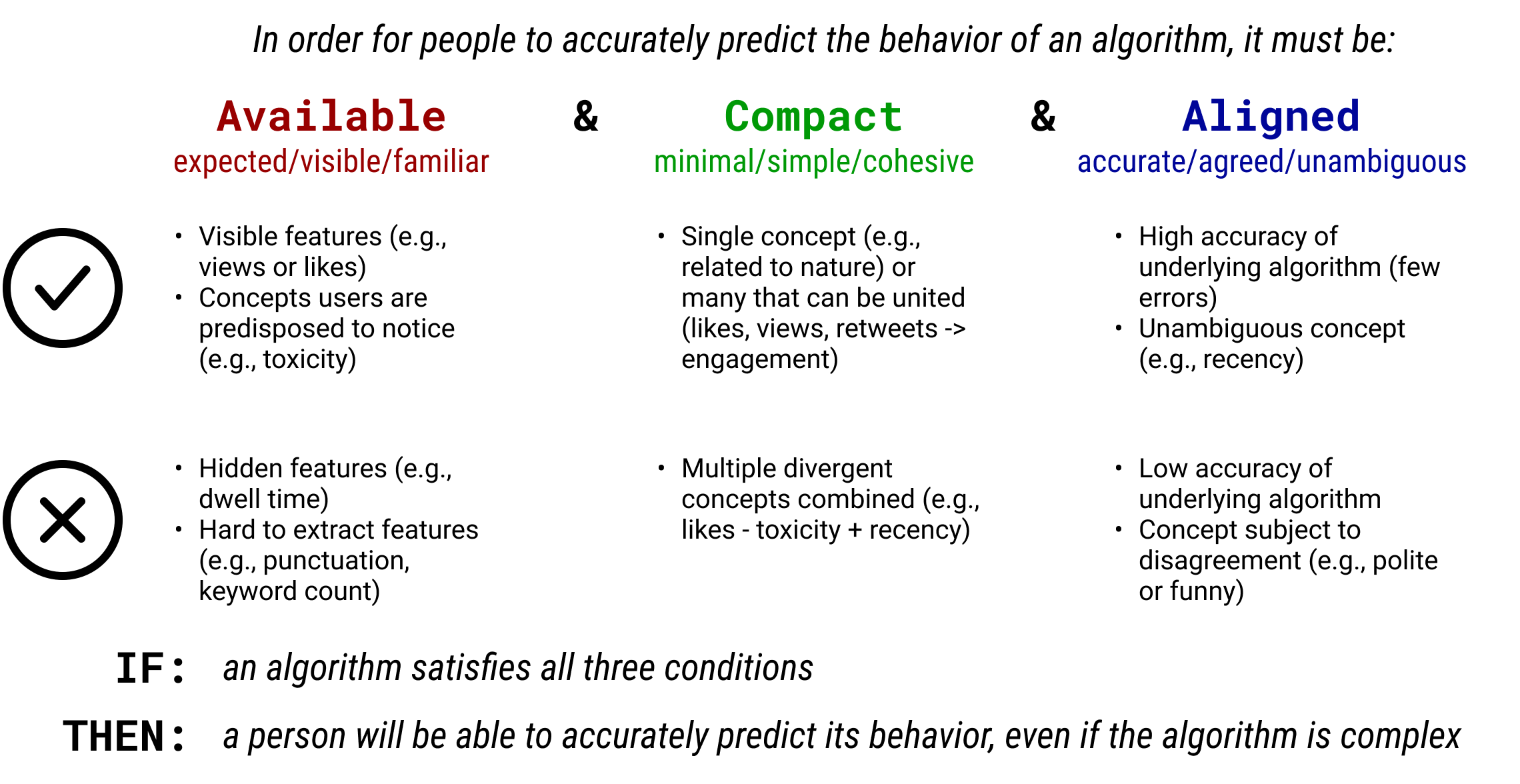}
    \Description[The core proposition: If an algorithm satisfies all three conditions, then a person will be able to accurately predict its behavior, even if the algorithm is complex. Characteristics improving and worsening each characteristic are shown.]{The core proposition: If an algorithm satisfies all three conditions (Available, Compact, Aligned), then a person will be able to accurately predict its behavior, even if the algorithm is complex. Characteristics improving and worsening each characteristic are shown. For available: visible features and concepts users are predisposed to notice help, hidden features and hard-to-extract features hurt. For compact: a single concept or many that can be united into one help, multiple divergent concepts combined hurts. For aligned, high accuracy of the underlying algorithm and unambiguous concepts help, low accuracy or concepts subject to disagreement hurt.}
  \caption{\textbf{ACA theory:} Algorithms satisfying the criteria of \textbf{A}vailability, \textbf{C}ompactness, and \textbf{A}lignment have the capacity to be predictable, even if the algorithm internal operation is opaque (e.g., in the case of complex deep learning models). These three criteria are mutually necessary in order for people to form accurate predictive mental models}
  \label{fig: theory}
\end{figure}

\subsubsection{Availability Construct: Recognition of System Affordances}
The first step in forming a mental model is recognizing the affordances that the system is conveying~\cite{norman2013design, norman2014some, johnson1983mental}. If someone cannot identify what signals the algorithm is using as inputs, or the signals it is displaying as outputs, none of the other stages can occur~\cite{norman2014some}. We propose the construct of \textit{availability}, defined as the cognitive availability~\cite{reber2016availability} of the algorithm's objective---how easily it comes to mind, influenced by salience and prior exposure---as a concrete criterion for algorithms that enables recognizability.

Availability draws from Tversky and Kahneman's work in cognitive psychology, specifically the availability heuristic~\cite{tversky1973availability}. The pair refer to availability as ``associative distance'', as in the cognitive effort required to leap from the current situation to the desired concept: ``the ease with which the relevant mental operation of retrieval, construction, or association can be carried out''~\cite{tversky1973availability}. 

In the algorithmic context, availability captures whether the algorithmic inputs and objective reflect concepts that the person would expect or leap to when seeing the algorithm's behavior. A social media feed algorithm that ranks by the number of bananas the post's author has eaten that day is not readily available, but one that ranks by upvotes is expected and thus available. Priming, through making concepts readier-to-mind, can thus help nudge availability. Likewise, a less available concept, for example the (very real) social media algorithm objective of showing you content that is predicted to produce replies for other users who otherwise have no feedback on their posts~\cite{eckles2016estimating}, is far less likely to lead users to produce a predictive mental model. However, availability is not inherent to the algorithm: availability is produced by the interaction between the algorithm, the user, and the context in which it is deployed. Engagement interactions, for example, would not be nearly so available as feed ranking features if they were not displayed on every post, while the more-likely-to-induce-comments algorithm could become more available if that prediction were conveyed visually. The content upon which the algorithm operates matters as well: an algorithm ranking the presence of sports topics in political tweets will not be particularly available due to both the lack of topical relevance (people would not be expecting such a ranking in that context) and due to the lack of contrast (most posts are likely to not concern athletics).

Advice that aids availability appears in prior human-AI interaction work, such as guidelines which espoused the importance of providing ``contextually relevant information,'' ``[making] clear what the system can do,'' and matching social norms so that the interaction ``is delivered in a way that users would expect''~\cite{amershi2019guidelines}. Feature importance explanations~\cite{lundberg2017unifiedapproachinterpretingmodel} can also be thought of as an availability intervention, by directing user attention toward the relevant features. In fact, some explainability work is focused on specifically making errors more available to users, by allowing them to more quickly or easily verify AI predictions~\cite{vasconcelos2023xai}. We emphasize availability beyond this feature-level to encompass broader patterns of behaviors that AIs have as well.

\subsubsection{Compactness Construct: Integrating Affordances into A Mental Model}
The second step of the mental model process is to distill and represent the affordances into a mental model~\cite{johnson1983mental}. Given those affordances, what principle of operation governs the system? Traditionally, mental models are referred to as capturing fairly mechanistic descriptions of system behavior (e.g., temperature controls on a refrigerator-freezer combo~\cite{norman2013design}), but this model can be constructed out of any concepts that the person's cognition treats as unified. For example, in a mental model of a snack vending machine, the selection buttons send a signal to rotate the metal spiral, which advances the bag of chips forward and out into the open air, and then gravity takes over and the bag falls into the opening of the vending machine. The manner in which electrical signals are sent is not directly represented in our mental model, but our cognition can still chunk it as a simple primitive that we use in our mental model. In fact, our chunking capabilities go beyond even this stage, and may cohere the behavior of the vending machine into simply, ``the item I select is there for me in the opening after I pay,'' without attention to how the vending machine achieves that transportation.

We capture this ability of our cognition---to chunk certain behaviors into simple primitives that we represent into mental models---via our second criterion, the construct of \textit{compactness}. 
We define algorithmic compactness as whether the algorithm reflects a unified concept, vs. a complex combination of diffuse concepts. Compact algorithms that reflect concepts the person is familiar with can be readily distilled and represented into a singular mental model. A simple combination of a small number of compact concepts can still remain compact, e.g., telling students that their exam grade will be determined by the number of standard deviations above or below the mean that they scored, or that Reddit's original algorithm was defined as upvotes minus downvotes.

Compactness anchors on the cognitive psychology concept of \textit{chunking}, the ``recoding of smaller units of information into larger, familiar units''~\cite{thalmann2019does}. Which concepts can get chunked together depend on expertise, as in a classic study of novice vs. expert chess players~\cite{chase1973perception}: expert chess players are more effective at reconstructing chess boards that they have only seen briefly, because experts build higher-level representations of the game than novices, who might attempt to memorize all the individual piece positions. A chunk unifies features together so long as they have stronger associations with each other than with other potential features~\cite{gobet2001chunking}. Likewise, an algorithmic objective that identifies content that is allowable in the United States under federal hate speech laws may be compact for a First Amendment expert, but not compact for lay users.

So, an algorithm is compact if it can be effectively chunked, or represented, into cognitive concepts. Simple algorithms are often compact by default: e.g., the decision criterion for a credit card might be a specific minimum credit score. For instance, stereotypical social media algorithms use a large number of underlying features, but for many people, those features can still be compactly represented under a single concept of ``engagement''. However, when that same algorithm adds multiple other concepts that cannot be represented (chunked) together effectively with engagement, such as political balance, implementation timeout limits on database servers, and predicted advertisement benefits, then the algorithm becomes less and less compact.

Underlying much interpretable AI work is the aim of reducing model complexity, which implicitly prizes compactness. Many of these projects aim to create simpler relationships between features or high level concepts involved in the model. For example, generalized additive models (GAMs) are designed to have simple associations between the conceptual building blocks of the model~\cite{hastie2017gam}, significantly reducing the complexity of the high-level algorithm behavior. Similarly, LIME explanations aim to create a compact model locally that users understand and can reason with~\cite{ribeiro2016lime}. Empirical work validates the importance of this criteria: providing a less compact representation (by revealing how the model actually worked) can hamper users' ability to predict model behavior~\cite{poursabzi2021manipulating}. Our definition differs by not focusing on the implementation-level (e.g., whether a model uses minimal features, or is clear-box, linear, or technically simple) but rather focusing on the behavioral-level (e.g., whether the model has behaviors that cohere to a whole).

\subsubsection{Alignment Construct: Agreement Between Our Application of the Mental Model and the Algorithm's Behavior}
The third and final step of the mental model process is for the person to simulate the mental model that they have built in order to predict the system's behavior~\cite{johnson1983mental, norman2013design, norman2014some}. Assuming the user has recognized and represented the algorithm's underlying concept, this simulation will produce an accurate prediction of an algorithm's behavior---but only if the person's simulation of that concept and the algorithm's execution of that concept produce the same results. If the two disagree---for example, if the person's mental model of toxicity disagrees with the algorithm's behavior---the person's predictions will be inaccurate. We capture this final criterion into whether or not the algorithm is \textit{aligned}: whether the user's understanding of a concept agrees with the algorithm's execution of that concept. 

Simple algorithms are often easily aligned: if we sort a social media feed reverse chronologically, the algorithm's implementation and the person's implementation are very likely to match unless the person makes a comparison error. Other algorithms are more difficult to align: while social media engagement ranking may be available and compact, alignment is often lower because the person cannot perfectly match the way that the algorithm weighs the factors in the engagement formula. A more subtle case involves when people themselves disagree: for instance, perceptions of humor, toxicity, or inappropriateness may also be misaligned for a given person if they disagree with the training data that the model was trained on~\cite{gordon2021disagreement}. 

Our intention behind the term ``alignment'' is most similar to the concept of ``representational alignment''~\cite{sucholutsky2024representationalalignment}, rather than alignment in the context of today's AI systems, which largely focuses on whether human values and goals are successfully embedded in AI systems. Alignment is a generalization of accuracy, which prior work argues is important for humans to understand AI behavior~\cite{2022-shared-interest}. In these contexts, accuracy is seen as both important for user understanding and in tension with interpretability via complexity~\cite{barredo2020xai}. Accuracy is also critical for behavioral measures of interest like user trust~\cite{rechkemmer2022confidence}. However, model accuracy alone is insufficient, we must also bridge the gap between user expectations and the model behavior~\cite{hong2020human, vafa2024llmhuman}. We emphasize this interpersonal difference in expectations for model behavior, where different users will have different internal understandings of the same concept that an algorithm might model. Our concept of alignment therefore broadens the scope of accuracy to consider when individual users would perceive a model as being accurate.

\subsubsection{Availability+Compactness+Alignment Allow Even Complex Algorithms To Be Accurately Predicted}

We aim to address the question of which algorithms people can understand. More specifically, we want to be able to classify whether people will be able to understand an algorithm well enough to predict its behavior accurately. We are also responding to claims in the literature that we must simplify algorithms in the social media context if we want people to be able to interpret them, reason about them, and ultimately work with them. We claim that even very complex algorithms---such as those including neural networks comprised of millions of parameters---can allow predictive mental models that enable interpretation, reasoning, and effective usage, if they satisfy all three of the constructs we have outlined above.

Our core proposition is that \textit{people can recognize, represent, and then reproduce an algorithm's behavior if and only if all three ACA criteria are met, no matter the algorithm's underlying complexity}. Our theory emphasizes two important contributions: first, that all three criteria are required, which clarifies why previous attempts at aiding predictability may have failed (by aiding one criteria but failing at least one of the others); second, that these three criteria are wholly independent of the complexity of the algorithm, opening opportunities for even extremely complex algorithms to enable predictively accurate mental models. 

Simple algorithms are often available, compact, and aligned. However, not always: \texttt{math.rand()} might be available and compact, but it is not aligned, so most people cannot predict its behavior. Likewise, an algorithm sorting texts by a term dictionary (e.g., the ``first person pronoun'' category in LIWC~\cite{pennebaker2001linguistic}) is highly compact and aligned once you realize what the algorithm is doing, but if the concept of first person pronouns is not available---not at top of mind so that you recognize it when you see it---it will fail the availability criterion and you will not be able to accurately predict the algorithm's behavior.  

However, critically, we claim that extremely complex algorithms can be ACA compliant and therefore predictable. Modern large language models can model concepts that are intuitive to people but very difficult to computationally model correctly. For example, computational social scientists have found that large language models can model constructs as accurately as experts across many domains~\cite{ziems2024can}. Likewise, we can rely on recursive convolutional networks to model dogs vs. cats in images correctly. Cats and dogs are highly available to us as concepts, they are compactly represented as basic categories~\cite{rosch1976basic}, and we agree with each other and with what is (at this point) a highly accurate algorithm. We argue, however, that if an algorithm fails even one criterion, people will be far less able to predict the algorithm's behavior: (Available:)~if an algorithm is trained to distinguish the American Crow and Common Raven, the concept is not highly available to many of us non-ornithologists, and so we cannot predict the algorithm's behavior; (Compact:)~if an algorithm is trained to flag if one of ten specific dog breeds is in the image but not any other dog breed, without any particular uniting factor for the ten, the algorithm is non-compact and would be difficult for a user to uncover in its entirety even if they were an expert on dog breeds; (Aligned:)~if an algorithm is trained to distinguish among dogs versus cats but is low accuracy and often makes mistakes, then the algorithm's execution is not aligned with ours, and so we cannot predict the algorithm's behavior. Importantly, our theory predicts that no other criteria are required; thus, ACA are not just necessary but also sufficient for peoples' mental models to be predictively accurate.

When all three ACA criteria are fulfilled, the high level behavior of the algorithm is itself an available, compact, and aligned mental model for the user. However, in cases where algorithms violate one or more ACA criteria, that option disappears, and users will not be able to form a predictively accurate mental model, therefore failing to accurately predict algorithm behavior. Much of the variety in users' theorization emerges from these scenarios where a fully satisfactory mental model does not exist and so they have to generate a model without the grounding of a ``good'' (e.g., ACA compliant) option. In these cases, users must compromise on the criteria to varying degrees, not necessarily mirroring the failure criteria of the algorithm itself.\footnote{We distinguish between the criteria that an algorithm fails and the criteria that a mental model fails. An algorithm failing one criteria does not entail that users will produce mental models failing that same criteria. Any algorithm, even those violating availability and compactness, can trivially produce a ACa mental model (available and compact but unaligned) by using something simple like chronological ordering that is just completely unpredictive of the algorithm behavior. However, users will struggle to produce a aCA mental model for any algorithm, by definition, because availability is key to being able to come up with the model at all.} 
Users are left making tradeoffs among the criteria to produce different mental models which satisfice in different ways, for example, using a broad and vague mental model which is available and compact but sacrifices alignment due to the lack of specificity.

A caveat: the criteria of being available, compact, and aligned represent ideal types~\cite{weber1925wirtschaft}, meaning that it is easy to conceptualize canonical examples and counterexamples, but actual algorithms will often live more on a spectrum. It may even be possible to construct a partial ordering---e.g., that one algorithm is more compact than another---but there is not necessarily a bright line that separates a compact algorithm from a non-compact algorithm. In the empirical section of this paper, for the sake of simplicity, we binarize each of these criteria and qualitatively classify algorithms as satisfying or not satisfying each criterion. We discuss possible measurements in Appendix~\ref{measures}.

Furthermore, algorithms are not globally available, compact, and aligned; rather, each factor may be true for a given individual. We describe algorithms as satisfying these criteria when they hold for the plurality of the population under consideration. In the case of our evaluation, the population is people in the U.S. who have some familiarity with social media. When practitioners deploy our theory, they should directly address the question of what knowledge and experience their target population has, affecting availability (through what they expect and have familiarity with), compactness (through which ``chunks'' they already intuit), and alignment (through their calibration on different constructs). While we may see interpersonal differences, they mostly stem from expert versus non-expert divides, in which case the designer should know which audience they are concerned with. Our study serves to demonstrate the applicability of this theory even when algorithms are classified on a broad population level.

In the following sections, we propose and describe an empirical study as a first test and falsification attempt of the theory we propose here. Theories can never be proven and only disproven, and ACA is no exception: we seek a strong initial test of its predictions to either demonstrate the predictive power of the theory or falsify it.
\section{Experimental Method}

We provide a first unified test of our theory by measuring people's ability to predict the behavior of algorithms that vary in ACA criteria. In this section, we describe our procedure for measuring the accuracy of algorithmic mental models via a prediction task. We describe our procedure, our process for selecting algorithm conditions, and our analysis method. Our experimental protocol was subject to oversight from our Institutional Review Board (Protocol \#71244).

The goal of this experiment is to test our theory by sampling a wide range of algorithms that span success or failure of our ACA criteria (Section~\ref{theory}) and measuring how well people can form predictive mental models of the algorithms' behavior. The theory claims that people will hold accurate predictive mental models if and only if all three criteria are fulfilled. So, we identify a range of algorithms that vary on their ACA criteria and test participants' abilities to predict those algorithms' behavior. Of course, as with any study, we cannot examine every possible current and future algorithm, so we purposely sample algorithms that cleanly reflect some combination of ACA criteria and are otherwise plausible\footnote{Some combinations of ACA criteria, such as failing all the conditions, intrinsically yield unrealistic algorithms because most useful algorithms will be at least one of compact or aligned---there is little point in being complicated if it doesn't produce higher performance. For this reason, some of the algorithm conditions are quite contrived.} choices for our study context. We then measure the relationship between satisfying these criteria and our participants' predictive ability for these algorithms. We pre-register our study at: \url{https://osf.io/gm5sr}. We discuss an earlier iteration of this study in Appendix~\ref{previous}.

These algorithms represent a variety of complexity levels, with some incredibly computationally simple---sorting by most likes, most recent, or most first person pronouns, which employ a simple count---and others using highly complex deep learning models---the severe toxicity algorithm based on the Perspective API, which is deployed for content moderation in real social media use cases~\cite{lees2022perspectiveapi}, or the GPT-based writing quality ranking. Other algorithms employ a combination of features, or algorithms of middling complexity, such as a BERT-based model that identifies expressions of caring and another BERT-based model that identifies expressions of conservative values.

\subsection{Main Task} \label{task_description}

We are mainly concerned with the accuracy of people's \textit{functional} mental models, or their ability to understand the input and output behavior of algorithmic systems, regardless of the true internal behavior~\cite{payne2007mental}. For this reason, we directly ask the participants to predict algorithm output based on input, a task demonstrating algorithm behavior understanding at a high level that maps to capability for desirable activities like self-advocacy~\cite{speith2024conceptualizing}. 

For our experiment, we require a task domain that allows for a wide range of different algorithmic objectives, enabling us to identify many algorithms that meet and fail each of the ACA criteria. We also require a domain that is widely familiar, and one that is now widely known as algorithmic. There are several possible candidates from the prior literature, for example AI decision tasks such as resume filtering and recidivism prediction. However, we sought a domain that intersects more with participants' everyday lives. So, we anchored our experiment in a social media context. Social media algorithms are omnipresent~\cite{narayanan2023}, and it is well known that they can vary in their objective: for example, a reverse chronological feed versus a feed that optimizes for engagement signals. 

In the context of social media feed algorithms, as well as many other algorithmic systems, users are not given a full specification of the algorithm but are rather exposed to its behavior over time, allowing them to develop folk theories of how it operates~\cite{eslami2016like}. We follow this same principle and expose each participant to the algorithm in the form of a ranked feed, where we show a collection of posts highly ranked by the algorithm at the top, and a selection of posts that are ranked very low at the bottom.\footnote{In the real life case, users would not see very low-ranked posts at the bottom of their feed, but instead could infer what they do not see through other use of the platform, discussions, and their intuition about what content exists. Given our limited time scale in the study, we inform participants upfront that the content that is ranked very low would be content they would be unlikely to see in a real social media setting.}  The posts contain information typically seen in social media such as Twitter/X or Bluesky, including likes, views, replies, retweets, post date and time, and the post text. The posts in the study are sampled from the home feeds of many different users on Twitter/X, and then filtered using GPT-4o to contain only posts with a political topic. We chose to focus on politics as a common topic that reduced the space of posts, allowing participants to more readily compare those that they see.

\begin{figure}[tb]
  \centering
    \includegraphics[width=0.9\textwidth]{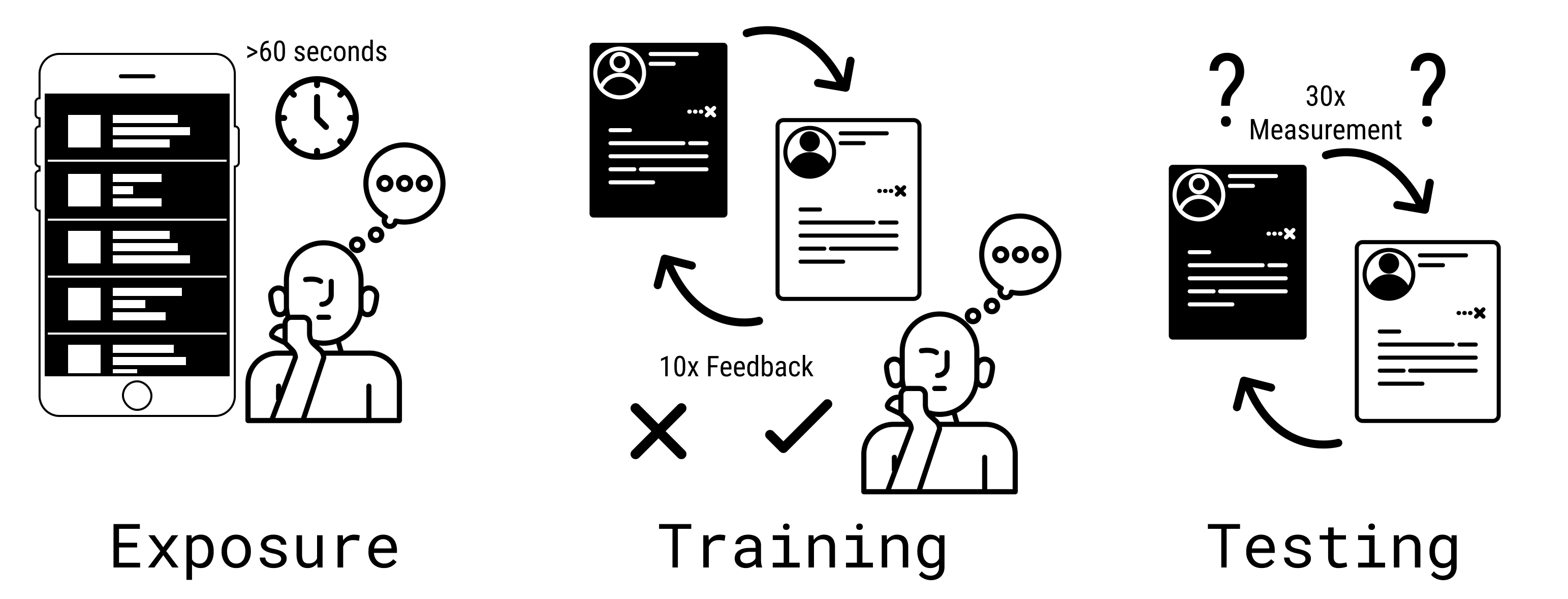}
    \Description[Visualization of three study phases, with 60 seconds spent on viewing the feed for exposure, 10 questions with feedback for training, and 30 questions without feedback for testing.]{Visualization of three study phases, with 60 seconds spent on viewing the feed for exposure, 10 questions with feedback for training (comparing two posts), and 30 questions without feedback for testing (also comparing two posts).}
  \caption{The process we use for measuring prediction accuracy takes place in three of the five study phases: exposure, training, and testing. In exposure, participants look at a particular feed and try to figure what the algorithm ranking it does. In training, they are given feedback on whether they correctly rank the pairs of posts they are given. In testing, they repeat the ranking task, but without feedback. }
  \label{fig: study process}
\end{figure}

Having seen the algorithm applied to a set of posts, but with no other information about its objective or functionality, the experimental prediction task is to select which of two posts the algorithm will rank higher. The system selects two posts from the dataset, then asks the participant to use their understanding of the algorithm they have just seen to predict which post would be higher in the feed order. The overall procedure is illustrated in Figure~\ref{fig: study process} and the exact user interface is shown in Figures~\ref{fig: study ui 1} and \ref{fig: study ui 2} in the Appendix.

\setlength{\tabcolsep}{6pt}
\renewcommand{\arraystretch}{1.3}
\begin{table}
    \centering
\footnotesize
\begin{tabular}{>{\raggedright}p{0.9in} c p{3.8in}}
Algorithm & Criteria & Description \\
\hline
Likes & ACA & Most likes \\
Reverse chronological & ACA & Most recent \\
Severe toxicity & ACA & Extreme toxicity (from Perspective API~\cite{lees2022perspectiveapi})\\
Factual & ACA & How much the content is presented as factual (rated by GPT 4o from 0 to 1 with 0.1 increments; prompts in appendix) \\
Quality & ACA & Writing quality (rated by GPT 4o from 0 to 1 with 0.1 increments) \\
Likes-Views-Retweets ratio & AcA & Likes minus twenty times retweets, divided by views \\
Split quality & AcA &
This algorithm alternates between ranking by writing quality and the inverse of writing quality depending on if the post has more than the median like count \\
Split engagement & AcA &
This algorithm alternates between ranking by likes minus twenty times retweets and twenty times retweets minus likes depending on if the post is above median toxicity \\
Engagement-Factual-Toxicity combination & AcA & A combination of views, retweets, and likes with toxicity and factual presentation of a post in a highly complicated way \\
First-person pronouns & aCA &
Highest first-person pronoun count \\
Odd digits & aCA &
Highest proportion of odd digits in the engagement features \\
Commas & aCA &
Highest proportion of commas relative to overall punctuation \\
Time-related & aCA &
How much the text of the post mentions time or timing (rated by GPT 4o from 0 to 1 with 0.1 increments) \\
Caring & ACa &
How caring a post is, using a BERT-based architecture that can report the presence of different values~\cite{kiesel2022identifyingvalues} that has medium performance on these posts \\
Achievement & ACa &
How achievement-focused a post is, using the same BERT-based architecture as the Caring algorithm \\
Noisy likes & ACa &
Most likes with noise added to the counts \\
Noisy factual & ACa &
Most factual framing with noise added \\
Grammatical features & acA &
10 times the ratio of second person to first person pronouns, plus the average word length, plus the number of punctuation marks \\
Split grammar & acA & Highest ratio of first person pronouns to all pronouns if the post has more than median likes, otherwise, by the highest ratio of commas to overall punctuation \\
Conservative values combination & Aca &
A combination of ratings for several concepts (tradition, achievement, personal security, and conformity to rules), performed by the same BERT-based model as in the Caring condition \\
Noisy likes-retweets-
views ratio & Aca &
Likes minus twenty times retweets, divided by views, with noise added \\
Noisy first-person pronouns & aCa & Highest number of first person pronouns, but with noise added \\
Wintertime & aCa &
How much the post relates to the concept of winter, specified unclearly (implemented with GPT 4o) \\
Noisy grammatical features & aca &
10 times the ratio of second person to first person pronouns, plus the average word length, plus the number of punctuation marks, with noise added \\
Caring-punctuation & aca &
The ranking of Caring added to the ratio of commas to overall punctuation
\end{tabular}
\medskip
\caption{We constructed a set of algorithmic conditions that varied the ACA criteria, with at least two algorithms for every combination of met criteria.}
\label{table: alg conditions} 

\end{table}

Our core measure is \textit{accuracy}: does the participant correctly select the post that the algorithm would rank higher? However, we also solicit information about users' mental models to complement our quantitative data.

In real-life interaction circumstances, people have the chance to form intuitions about the algorithm behavior and then see them confirmed or denied by further exposure. Since our study is not longitudinal, we proxy this experience with a series of training questions; after each training question, participants receive feedback about whether their predictions are correct or not. Further information about study procedure is found in Appendix~\ref{procedure}.

\subsection{Conditions}

We adopt a fully between-subjects design, assigning participants to one of 25 algorithmic conditions described below. We include multiple algorithms for every possible permutation of ACA criteria. We shorthand the criteria met by the algorithm: a capital letter to signify that they meet that criterion, and a lowercase letter to signify that they do not meet that criterion. For example, an ``ACA'' algorithm is available, compact and aligned; whereas an ``acA'' algorithm is not available nor compact but it is aligned. The conditions are defined within Table~\ref{table: alg conditions} and their assignments are justified in Appendix~\ref{condition_justification}.

\subsection{Hypothesis}

$\mathcal{H}$: In accordance with our theory, we hypothesize that algorithms fulfilling all three criteria--Availability, Compactness, and Alignment--will correspond to higher predictive performance for our participants than algorithms that fail to fulfill at least one of the criteria.

\subsection{Participants}

We recruited a total of 1250 participants to participate in this study, averaging 50 per condition. We chose our minimum sample size per ACA category (different combinations of availability/compactness/alignment; e.g., AcA, acA, Aca, each of which have multiple algorithms) based on a power analysis done on pilot data and powered at 0.95, with $\alpha = 0.05$, to detect an effect of at least 10\% over the 50\% random guess baseline. That power analysis determined that we should have at least 100 participants per category. Each category has a minimum of two algorithms, so we set our sample size to 50 per algorithm, to keep our algorithm-level sampling consistent. 

Our participants are recruited from Prolific (\url{prolific.co}) and CloudResearch Connect (\url{connect.cloudresearch.com}), two online crowdsourcing platforms. 
We recruited participants that had at least 50 submissions, were located in the United States, were native English speakers, had an approval rating of at least 95\%, and that had not completed our task before. All participants received a payment of \$5 for 20 minutes of their time with bonuses of \$0.10 for every question they got correct over a 50\% accuracy rate (15/30).

\subsection{Analysis}

In line with our hypothesis, we aim to statistically test the relationship between an algorithm's satisfaction of the ACA characteristics---availability, compactness, and alignment---and people's ability to predict said algorithm's behavior. We therefore analyze our results with a mixed effects logistic regression of specification: \begin{verbatim}
    accurate ~ available * compact * aligned + (1 | participant) + (1 | platform)
\end{verbatim}
This model specification is designed to test our hypothesis that all three criteria are needed, thus the three-way interaction.\footnote{Note that the syntax a * b * c is a convenience syntax that includes (1)~the three-way interaction abc, (2)~the multiple pairwise interactions ab, bc, and ac, and (3)~the individual main effects a, b, and c.} The three criteria--available, compact, and aligned--are coded as binary variables. We include random effects for participant ID and platform (Prolific versus Cloud Research Connect).

We then perform an additional, follow-up Tukey test to distinguish differences in prediction performance for algorithms satisfying all three ACA criteria.

We also analyze the textual data from the free-response questions about participant mental models. We code participant mental model specifications in multiple ways: deductively coding the degree of the mental model's adherence to the true algorithm specification and the number of features, and inductively coding the content of the mental model to observe patterns in types of features used. For the degree of mental model matching to the underlying algorithm behavior, we used a scale from 0 to 3, with 3 representing a complete match, 2 a strong match, 1 a weak match, and 0 representing no real matching. Two authors coded all participant responses, with strong inter-rater reliability (linear weighted $\kappa=0.85$, Pearson $\rho=0.91$). For the number of features, two authors overlapped on coding 10\% of the data, with substantial inter-rater reliability (linear weighted $\kappa=0.80$, Pearson $\rho=0.86$).

Once an initial code set of mental model features is established, we use GPT-4o to code for inclusion or exclusion of certain features (engagement, time, partisanship, and controversial framing). The accuracy of the GPT-coding is verified by the authors comparing hand-coded results for 100 participants; results were only used after there were no errors found, corresponding to $\kappa$ of 1. All other coding is done by hand, by members of the author team. We also (deductively) code the participants' descriptions of their process to identify when they solidified their mental models and how they changed.

\section{Results} \label{results}

\begin{figure}[tb]
  \centering
    \includegraphics[width=0.9\textwidth]{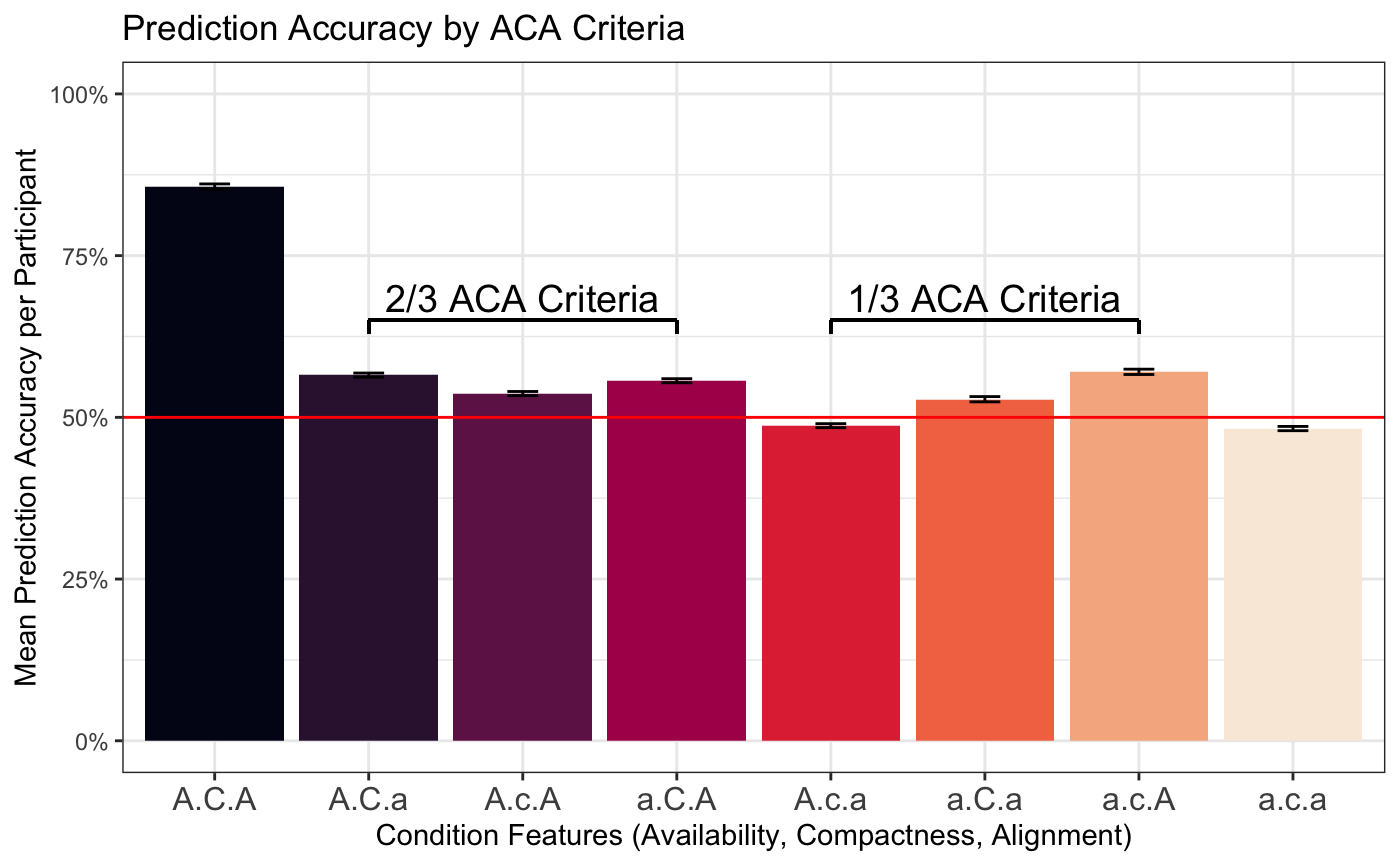}
    \Description[A bar plot of prediction accuracy by ACA criteria. The bar for ACA algorithms is around 85\%, while the bars for all other configurations are close to 50\%.]{A bar plot of prediction accuracy by ACA criteria. The bar for ACA algorithms is around 85\%, while the bars for all other configurations range around 45-55\%. There is no apparent difference for 2/3 ACA criteria versus 1/3 or 0/3 ACA criteria.}
  \caption{Participants predicted algorithm behavior with the highest accuracy for algorithm conditions that satisfied all three ACA criteria. Most algorithms that failed one or more criteria had predictive performance at close to baseline rates.}
  \label{fig: s1 barplot}
\end{figure}

In this section, we first address our hypothesis through our statistical results.
We visualize our results in Figure~\ref{fig: s1 barplot} and include the model output in Table~\ref{table: study 1 results}. Then, we describe results regarding the process of mental model creation and deployment, to explain how our theory predicts the mechanism (not just result) of mental model usage. 

\begin{table}[!htbp] \centering \small
\begin{tabular}{@{\extracolsep{5pt}}lr} 
 \multicolumn{2}{c}{\underline{    Mixed-effect Logistic Regression    }} \\ 
 & \textit{Dependent variable: Accuracy} \\ 
\hline 
 Intercept & $-$0.069 \hspace{0.2cm} (0.185)\\ 
 platform & $-$0.046 \hspace{0.2cm} (0.045) \\ 
 availability & 0.044 \hspace{0.2cm} (0.260) \\ 
 compactness & 0.177 \hspace{0.2cm} (0.262)\\ 
 alignment & 0.394 \hspace{0.2cm} (0.257)\\ 
 availability:compactness & 0.166 \hspace{0.2cm} (0.346)\\ 
 availability:alignment & $-$0.233 \hspace{0.2cm} (0.341)\\ 
 compactness:alignment & $-$0.182 \hspace{0.2cm} (0.343)\\ 
 \textbf{availability:compactness:alignment} & \textbf{1.826$^{***}$} \hspace{0.2cm} \textbf{(0.447)}\\ 
\hline 
\hline \\[-1.8ex] 
\end{tabular}
\medskip
\caption{In the mixed-effect logistic regression, we see a significant positive coefficient for the ACA interaction alone. This indicates that having all three criteria was the most advantageous for predictive ability. All other interactions between ACA criteria do not show significance. Note: $^{*}$p$<$0.05; $^{**}$p$<$0.01; $^{***}$p$<$0.001. Standard errors are included in parentheses.}
\label{table: study 1 results}
\end{table}

\subsection{People can accurately predict behavior of algorithms that are available, compact, and aligned}
Participants performed more accurate predictions for algorithms fulfilling all three conditions, with performance dropping when even one criteria was not fulfilled: the average accuracy in participant prediction for an ACA algorithm was 85\% versus 54\% for a non-ACA algorithm. Our statistical model validates this result: the three-way interaction effect of availability, compactness, and alignment in our mixed effect logistic regression has a statistically significant positive coefficient. The coefficient (log odds) is 1.826, which signifies an odds ratio of 6.2 for correct versus incorrect predictions when satisfying all three criteria versus none. None of the other coefficients yield statistical significance, which means there is no evidence indicating that any of the ACA criteria individually produced more accurate prediction, nor any of the pairwise combinations of criteria---only when all three were simultaneously satisfied, supporting $\mathcal{H}$. 

\begin{figure}[tb]
  \centering
    \includegraphics[width=\textwidth]{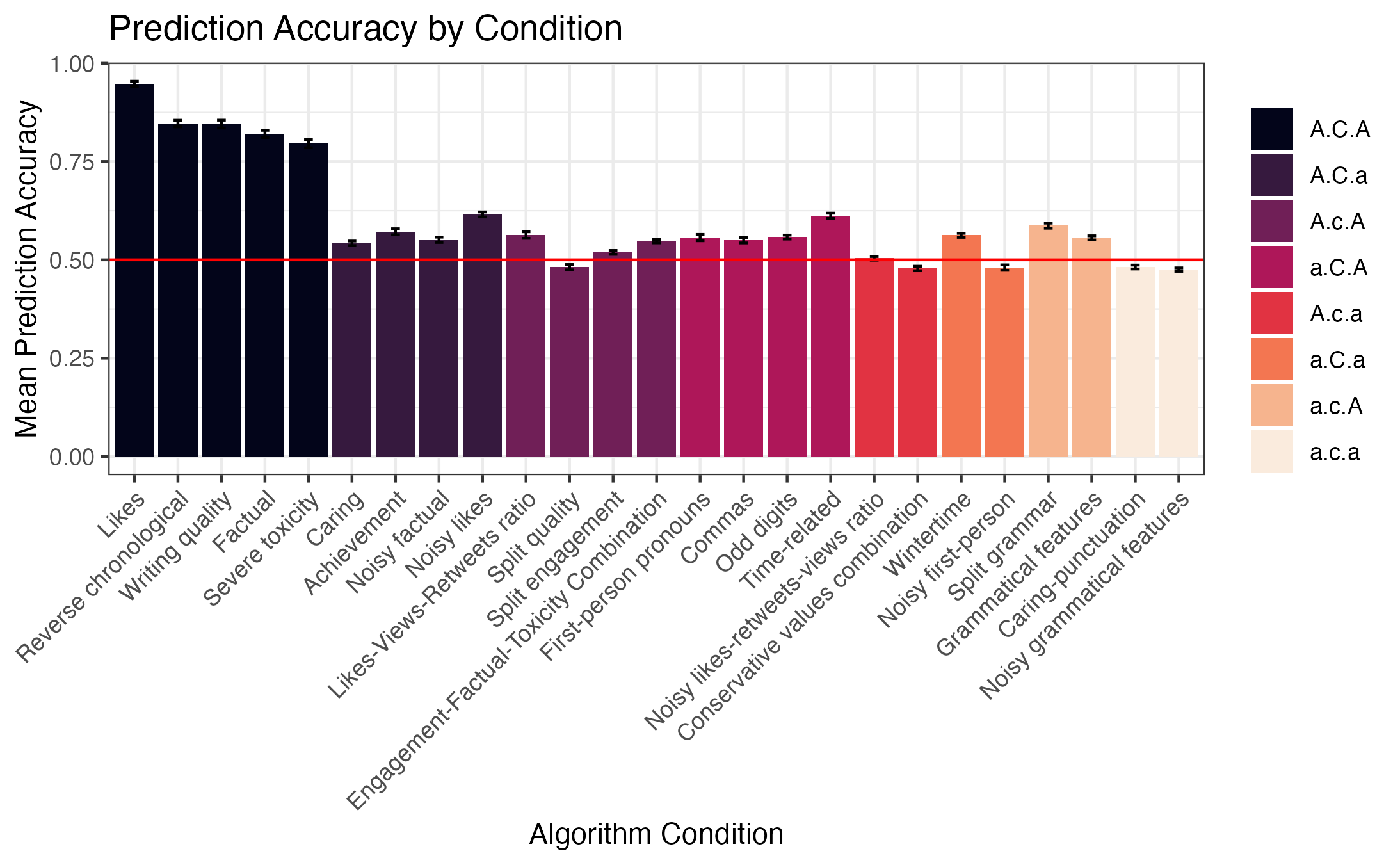}
    \Description[A bar plot of prediction accuracy across all algorithm conditions. Likes is notably the highest, and all ACA algorithms are above 75\%, while the highest of the non-ACA algorithms are around 60\% and most are closer to 50\%.]{A bar plot of prediction accuracy across all algorithm conditions. Likes is notably the highest, at around 90\%, and all ACA algorithms are above 75\%, while the highest of the non-ACA algorithms are around 60\% (Noisy likes and Time-related) and most are closer to 50\%.}
  \caption{Participants predicted algorithm behavior at close to 80\% or above for algorithms that satisfied all three ACA criteria. Most algorithms that failed one or more criteria had predictive performance within 10\% of a random guessing baseline (50\%) rate. Only two non-ACA conditions (noisy likes and time-related) had prediction accuracy at slightly above 60\%.}
  \label{fig: s1 barplot condition}
\end{figure}

We do see variance across algorithm conditions, with some non-ACA algorithms experiencing prediction performance around 60\%. However, there remains a large gap between the highest performing on the non-ACA algorithms (61.6\% accuracy) and the lowest performing of the ACA-satisfying algorithms (79.6\%), as seen in Figure~\ref{fig: s1 barplot condition}. 

\subsection{Accurate mental models correspond to accurate behavior predictions}

While our hypothesis was constructed with the underlying assumption that people's mental models would match up with the algorithm's actual objective, many participants deployed completely different mental models to varying levels of success. 

\begin{figure}[bt]
  \centering
    \includegraphics[width=\textwidth]{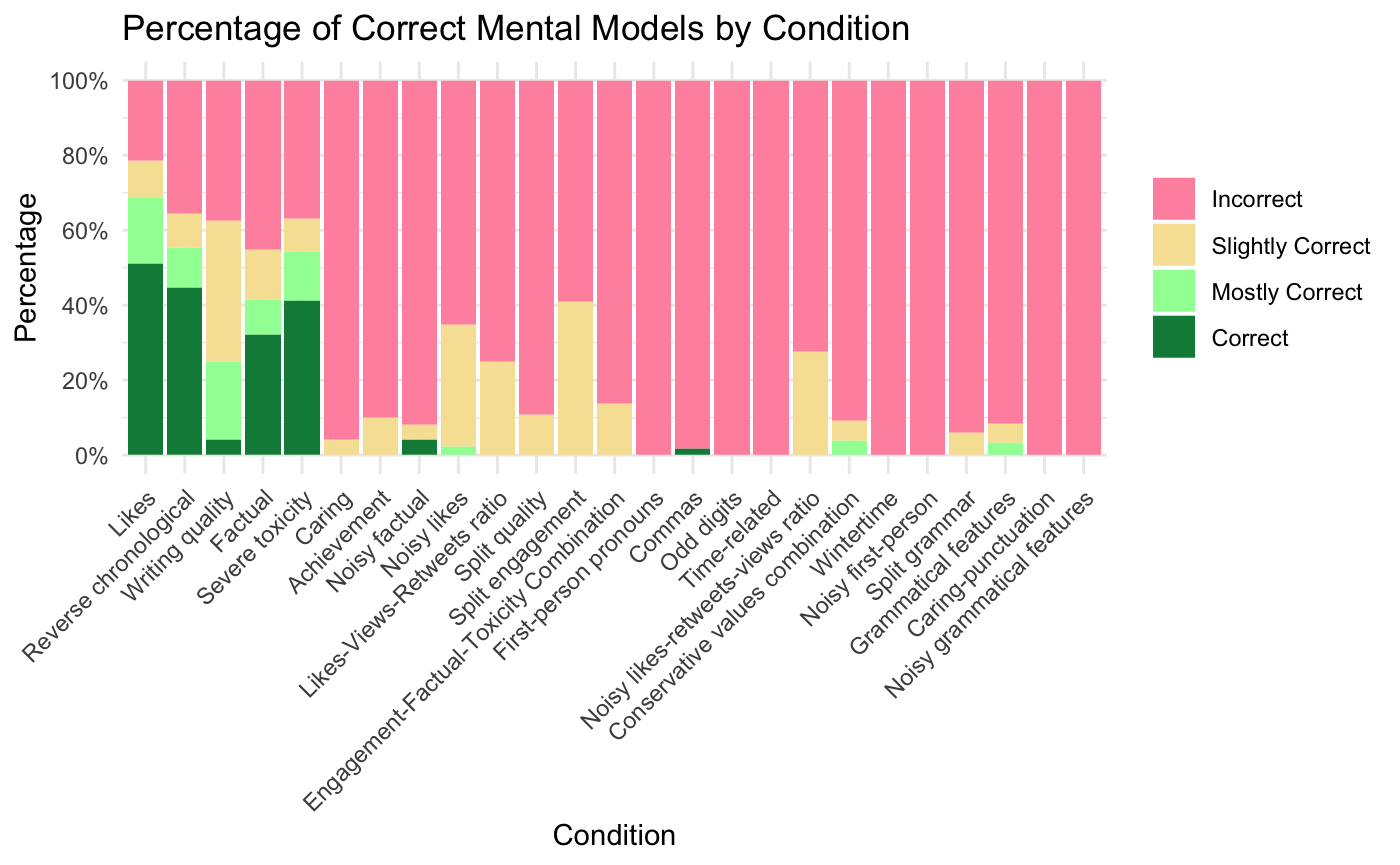}
    \Description[A plot showing the distribution of correctness across different algorithm conditions. Every non-ACA algorithm has majority completely incorrect mental models. The ACA algorithms all have at least half slightly correct or more.]{A plot showing the distribution of correctness across different algorithm conditions. Every non-ACA algorithm has majority completely incorrect mental models. The ACA algorithms all have at least half slightly correct or more. Likes, reverse chronological, factual, and severe toxicity have high levels of completely correct, while writing quality has notably lower high correctness.}
  \caption{We visualize the percentage of mental models for each algorithm condition that matched the design of the algorithm. The ACA conditions exhibited higher levels of matching, though the writing quality condition had an unusually low level of complete matching compared to the other ACA conditions. }
  \label{fig: mm matching}
\end{figure}

Average prediction accuracy monotonically increases along with the degree of mental model agreement. The average prediction accuracy for participants with a completely agreeing mental model (rating of 3/3) was 97.5\%, and for a high match (rating of 2/3) was 90\%. This result suggests that the process of \textit{applying} a correct mental model was rarely a stumbling block for participants. However, the majority of participants across all conditions except Likes used alternate mental models, which indicates that the process of recognizing or synthesizing the mental model was a bigger hurdle. In fact, only 133 (out of 1250, or 11\%) participants had a highly to completely matching mental model, 125\footnote{A few participants correctly guessed unavailable algorithms, such as ranking by comma prevalence, or unaligned algorithms, such as ranking by likes with noise added.} of whom were in ACA algorithm conditions and only 8 in non ACA conditions. In other words, 61\% of participants in ACA conditions had a highly or completely matching mental model, while less than 1\% of those in non-ACA conditions did. The degree of mental model matching across condition is visualized in Figure~\ref{fig: mm matching}.

At the same time, many participants managed high prediction accuracy with partial or no mental model match. 8\% of participants had prediction performance of 80\% or above while using a mental model with low or no agreement to the true algorithm objective (17\% above 70\%). In following sections, we look into how and why participants deployed these seemingly ``incorrect'' mental models, and how they traded off the three different ACA criteria to make their selection.

\subsection{Participants can leverage correlations to achieve accuracy even with incorrect mental models}

Many algorithm conditions had very strong correlations with other potential algorithms (or features), allowing participants to achieve accuracy in their predictions while using technically incorrect mental models. The most obvious example of this is for the likes algorithm, which only sorts by likes. However, participants commonly voiced 
alternate mental models involving retweets, replies, and views (and saw high prediction accuracy), an unsurprising misconception and result when the amount of retweets, replies, and views all have strong correlations with the number of likes.

\setlength{\tabcolsep}{6pt}
\renewcommand{\arraystretch}{1.75}

\begin{table}[tbh]
\centering
\small
\begin{tabular}{llc}
\textbf{Main Algorithm} & Matching Algorithm  & Percent Ranking Agreement\\
\hline
\textbf{Likes} & Replies & 100 \\
 & Retweets & 100 \\
 & Views & 100 \\
\textbf{Chronological} & Right-wing & 72\\
 & Likes & 70\\
 & Views & 71\\
\textbf{Factual} & Non-toxicity & 77\\
\textbf{Writing quality} & Left-wing & 71 \\
 & Non-toxicity & 85 \\
\textbf{Achievement} & Non-toxicity & 76 \\
\textbf{Likes-retweets-views ratio} & Right-wing & 79 \\
 & Replies & 76 \\
 & Retweets & 82 \\
    \end{tabular}
    \caption{Many of our algorithm conditions had high correlations with highly available features or concepts. We measure the degree to which two algorithms align by calculating the percent of pairwise rankings which remain the same across algorithms. Note that this measure is not commutative, because the posts used in the measurement change slightly depending on which algorithm is first (they are chosen to be extremely low or extremely high-ranked by the main algorithm). This table contains only algorithm pairs that agree at a rate of 70\% or above. }
    \label{table: alg correlation}
\end{table}

This pattern persisted across other conditions, though with less intuitive algorithm correlations perhaps due to underlying patterns in our social media posts. For example, one participant explained ``I basically just put the post with the lower word count at the top and more often than not I was always proven right for doing so,'' and managed 87\% accuracy, despite her condition being the likes-retweets-views ratio algorithm. Similarly, two participants in the first person pronouns condition had near-perfect prediction accuracy across practice and test questions while using a mental model which ``ranked any posts that were 15, 16, or 17 days old lower.'' Beyond these lucky few cases with this near-perfect agreement, many algorithms had a lower level of underlying correlation with readily available features such as likes, views, time posted, partisan leaning, and toxicity. We report our different algorithm conditions alongside all available features or concepts for which rankings agree\footnote{Since our ranking task involves comparing very differently ranked posts, we evaluate the similarity in two algorithms not through simple correlation but through comparing how many post rankings would be the same.} at least 70\% of the time in Table~\ref{table: alg correlation}.

\subsection{Participants tend towards available and often compact mental models, even when incorrect}

\begin{figure}[tb]
  \centering
    \includegraphics[width=\textwidth]{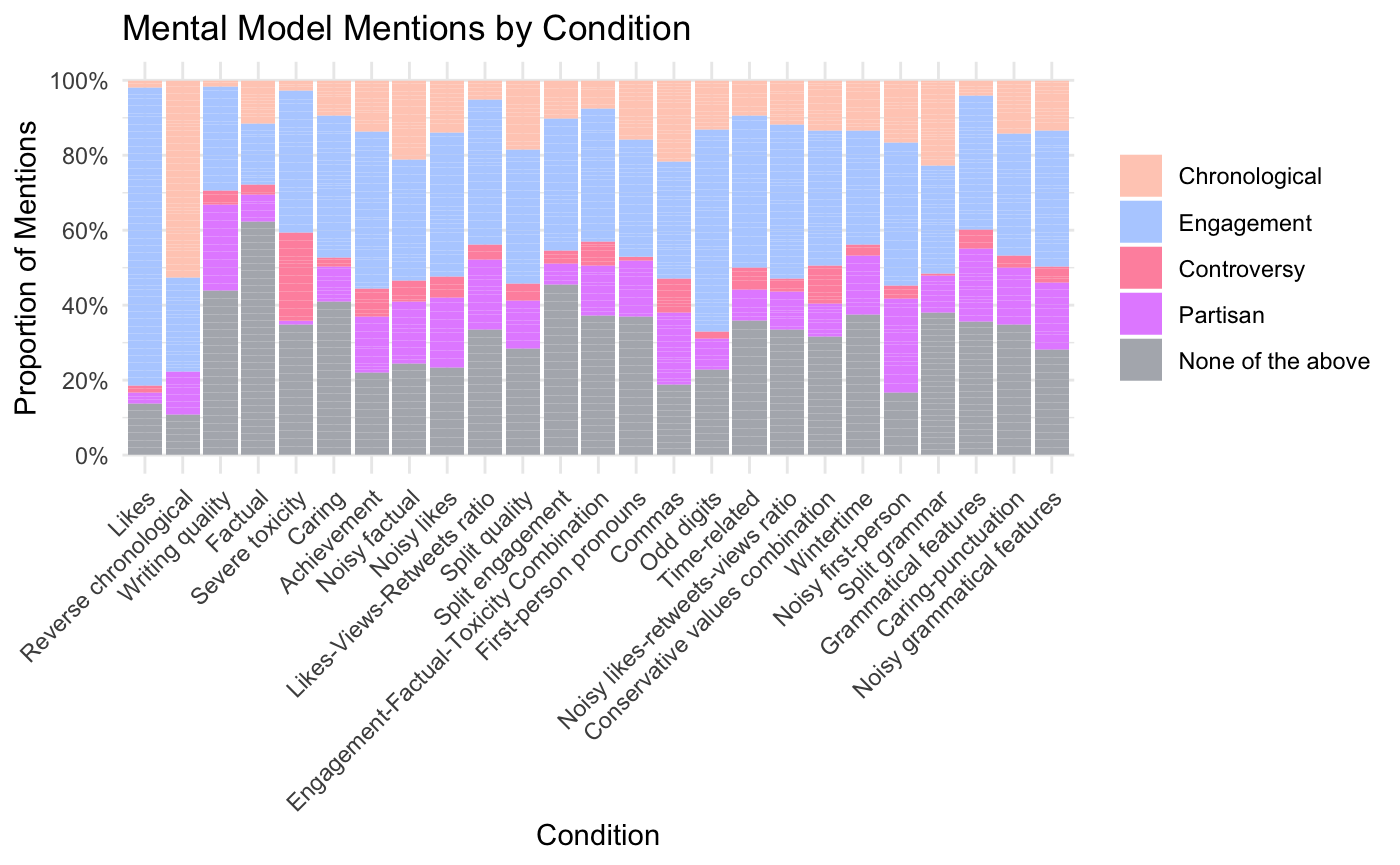}
    \Description[A plot of the distribution of common mental model features mentioned across algorithms. Engagement features are commonly mentioned across all algorithms, and temporal, partisan, and controversy-related features are also mentioned frequently.]{A plot of the distribution of common mental model features mentioned across algorithms. Engagement features are commonly mentioned across all algorithms, especially for the likes algorithm, and temporal, partisan, and controversy-related features are also mentioned frequently. Inclusion of none of these features ranges from around 10\%-60\% depending on algorithm.}
  \caption{The percentage of mentions, normalized per participant, of each of the major mental model feature categories. Engagement was frequently mentioned, regardless of its true relevance to the category, emphasizing the role of availability and priors in peoples' construction of mental models. }
  \label{fig: s1 distribution}
\end{figure}

When participants did not enlist a correct mental model, they overwhelmingly theorized algorithms that used highly available concepts or features. These can be clustered into two main categories: those available for social media broadly, and those available in the political domain. In category one, we see features such as likes, reposts, replies, and views, as well as time posted, plus higher level concepts like controversy and toxicity. In the latter category, we see participants predisposed to notice patterns of partisan leaning or political topic. 44\% of participants in non-engagement-based algorithm conditions believed the algorithm involve engagement-related features when it did not, 19\% believed the same for posting time, 17\% for partisan leaning, and 6\% for controversy. We visualize the distribution of these features in mental models across the different algorithm conditions in Figure~\ref{fig: s1 distribution}.
In fact, participants were so predisposed to consider these features (especially those related to engagement and chronology) that they would even mention having ruled those out in many cases where they did not play a role in the mental model, e.g., ``Dates and likes did not seem to impact how they were ranked.'' 

Participants also tended towards compact mental models, though to a lesser degree than they did for available algorithms. A slight majority of participants had a mental model with a single feature (652, or 52\%). Even among incorrect mental models, almost half (49\%) were a single feature, despite the fact that many mental models were incorrect specifically due to the addition of extra, unrelated features. There was a decaying pattern across number of features in the mental model, with 652 single-feature mental models, 339 two-feature, 130 three-feature, and so on.\footnote{Some (85) participants are coded as having no features in their mental model due to them stating that they have no idea what the algorithm behavior was.}

However, when participants enlisted these alternate mental models, they achieved only middling levels of prediction accuracy. The average prediction accuracy for a completely incorrect mental model was 55\%, with little variation in accuracy according to number of features. In these cases, they often had little faith in their predictions, like one participant who described, ``I'm really not sure. I think I seemed to see a pattern of more right leaning topics being shared on the top, but it was hard to tell sometimes.'' This sort of mental model description, with statements of uncertainty, was very common, with other participants hedging their predictions similarly with phrasings like ``I'm still not sure,'' ``I think the algorithm was based off X, but maybe Y,'' and ``it was ranked by X and Y? idk.''
We therefore see a tendency among participants to satisfice by picking an available and often compact but only partially aligned mental model when a fully ACA one is not available. 

\subsection{Participants produce fuzzy and non-compact mental models to maintain plausible correctness in the face of uncertainty}

Not all participants chose a compact mental model. Our theory might suggest that participants should gravitate towards ACA alternate mental models when the underlying algorithm behavior does not meet these criteria. However, sometimes the available and compact alternate mental models are not sufficiently aligned for a particular algorithm. In these cases, participants accepted a less compact state in order to account for the errors that the nearest available and compact algorithm could not explain.

We can see this tendency through the numerous ($N>50$) participants who specifically describe needing to adjust their mental models during practice to include more features after their initial, more compact, mental model was disproven by a practice question. For example, one participant described that, ``It was the time posted at first but then the practice threw me off when I got a few wrong. I realized it was also the engagements as well. I thought it maybe included the texts which were political.'' While their initial mental model was available and compact, they added new features after getting practice questions wrong, eventually ending up believing that the algorithm, "...put the newer ones on top which is according to the time posted. Then it based on the engagement which was replies and retweet depending on how many replies there were and retweets as well as likes.'' 

Overwhelmingly, participants who described non-compact mental models did not specify them completely, describing the algorithm behavior only in terms of the different features and not how they would be combined to actually deploy. For example, one participant described that, ``I think [the algorithm] had something to do with views, retweets, date posted and if it was about trump and or biden.'' This imprecise description allows the mental models to exist in an immature state where they are not fully usable but difficult to disprove---some combination of the described features would likely explain the visible behavior, but said combination is hard to discover. 

\subsection{Conditions exhibit minor differences beyond binary ACA factors}

Comparing within the ACA-compliant algorithms, we see some statistically significant differences between conditions, with post-hoc Tukey tests revealing that the only pairwise significant differences between conditions were between the Likes algorithm and all others ($p<.0001$). Given that we also see many participants incorrectly attributing algorithm behavior to engagement features, we could theorize that Likes is a particularly available algorithm that makes it more likely to be deployed. 
From this example, we can see that the binary ACA classification does not explain \textit{all} of the variation in prediction accuracy. 

We also notice some variation along the non-ACA conditions that reflects gradation in our criteria. For example, the Noisy likes algorithm and the Time-related algorithms both have prediction accuracy above 60\%, quite above baseline performance. While latent correlations of varying availability and alignment could explain some of this improved performance, it is more likely that Noisy likes is more aligned than Noisy factual because likes itself is more available and has higher accuracy (being explicitly accurate to its underlying concept) than the factual algorithm.

\section{Discussion}

This paper proposes that algorithms capturing mental models that are available, compact, and aligned can yield accurate predictions from people. Here, we discuss the limitations of our experimental approach alongside suggestions for future work, practical considerations for applying our theory, and implications for algorithm design in the social media context and beyond.

\subsection{Limitations and Boundary Conditions}

Our study supported the ACA theory we propose but future work should test it further and attempt to identify systematic holes. We suggest three directions that remain unaddressed by our experiments.

We focused on a set of algorithms in a single context: social media feed ranking algorithms. Testing ACA in additional domains is an important avenue for future work because users have different needs and intentions in different cases and thus the criteria determining their appropriate understanding could differ as well. 

We additionally limited our experiment to political posts alone, not the full range of topics found on social media. This likely made algorithms involving text semantics (e.g., toxicity) easier to predict, because the posts had more in common with each other, making their differences more stark. With a more ecologically valid, diverse feed, we would expect to see the same split between ACA algorithms and non-ACA algorithms, but with more noise, coming from users considering the topical differences as a potential signal.

We did not use many actual ranking algorithms in use by platforms. Our theory would predict that many such algorithms suffer from lower prediction accuracy due to non-compactness, which aligns with qualitative user research depicting confusion on platforms like TikTok and Facebook~\cite{eslami2016like,karizat2021algorithmic}. We did not broach personalized algorithms, either. While we believe that personalized recommendations may be highly aligned and available for people (e.g., ``it picks what I like''), our studies do not address this. 

Our study was a single shot online experiment, with attendant limits to generalization. Users normally form folk theories of social media algorithms over days, weeks, or months. Our work examines what participants can glean in a limited time-frame without assistance. Also, while users of social media platforms often experience the algorithm changing over time, we provided a static algorithm, so our study does not take into account possible factors affecting users when they adapt their mental models in response to changes.

We focused on the idea of algorithmic behavior---what will the algorithm do when given these inputs?---which can often be abstracted away from the internal algorithm operations for highly-performant algorithms. However, other understanding tasks, such as error prediction or explanation, have different requirements. Often, detecting errors requires an understanding of the internals, so users can know what the algorithm is doing when it does not achieve its aims. Therefore, our theory would need to be re-interpreted and extended in order to cover these additional cases. For example, if a user only uses social media to casually browse, looking for entertainment, the mental model of ``sorting by engagement, with inappropriate content down-ranked'' may suffice. However, if a user is posting about highly sensitive topics like past self-harm and seeking connection with others in recovery, the exact distinction between appropriate and inappropriate content may be extremely important for their mental model.

\subsubsection{Directions for Future Work}

In order to validate this theory more broadly, further experiments should probe the explanatory power of availability, compactness, and alignment in more domains and task contexts. Domains like hiring, search, and insurance are worth studying to test the generalization of ACA. We also recommend introducing personalization into algorithm conditions, due to its increasing ubiquity in social media algorithms and other contexts. Finally, we believe future work should explore different task conditions tailored to different types of understanding. For example, tasks could be derived from different common user intentions or concerns (e.g., Can they interact with the feed in a way that will impact their feed in a particular way? Can they change a post in order to get it ranked higher? Can they tell if the algorithm involves a particular characteristic or not?) to reflect how well their algorithm understanding serves those specific informational goals.

\subsection{Nuances in Applying ACA Theory}

While we choose algorithms that cleanly meet or fail the ACA criteria, classifying or measuring in more ecologically valid scenarios may be difficult. In fact, we believe there is significant nuance in understanding and applying this theory, worth discussing here.

Firstly, availability can change over time. What starts out as unavailable may become better understood with time, as processes like algorithmic gossip~\cite{bishop2019gossip} help disseminate theories about algorithm behavior; everyday users are known to rely heavily on exogenous information in their folk theorization efforts~\cite{devito2021adaptive, devito2018form}. However, designers should not count on noisy and slow learning processes to overcome unclear and confusing algorithm deployment; if understanding behavior is important, it is likely important from the beginning and should be facilitated by the designer \textit{at that point}. Additional factors like the actual posts or the design of the algorithm interface can aid or hinder availability: the presence of strong difference along a particular variable is necessary to recognize a ranking based on that variable, and a stark contrast can make the ranking stand out. 

The division line between sufficiently and insufficiently compact algorithms is unclear. While a single variable is obviously compact, and a hundred combined non-cohesively is not, there is not an obvious inflection point in feature count where an algorithm becomes too complicated to comprehend. When a certain combination of factors becomes commonplace, it can become a single unit, understood for itself and not as the sum of its parts. Likewise, as students proceed through a university education, they build new compact concepts as they learn, much like a chess expert.  

Finally, alignment appears from our study to be the factor with the greatest leniency. Participants were quite robust to errors, and in fact regularly deployed incorrect mental models with around 70\% accuracy. This tendency indicates that not only are patterns still recognizable at that level of noise, but also that people will accept these patterns as representative.

The impact of alignment did not always show at the stage expected (e.g., low alignment resulting in poor accuracy for the mental model application). In fact, all criteria were most impactful in terms of gating initial selection of a matching mental model. The vast majority of participants succeeded in their predictions if they could specify the correct mental model. All three criteria affect a person's ability to find and \textit{validate} an appropriate concept (or set of concepts) from which to produce their mental model. Even alignment matters for selecting the mental model because one that cannot be deployed accurately will not be picked: checking is half of the guess and check process. One participant explicitly referenced dropping a mental model when it did not check out: ``I thought at first it was ranked by news versus opinion but then I had no idea.'' Since they were assigned to the noisy factual algorithm, which specifically added errors to an algorithm meant to rank news-like or factual content above opinion content, this participant shows how failing alignment can cause a participant to not use a particular mental model when it was otherwise available and compact. In this way, alignment is still causing a failure in applying a mental model, this application step is just done prior to the test questions.

On a user-by-user basis, compactness and alignment may differ significantly. For example, certain algorithms may be compact for people due to their domain knowledge: a physician may find a cluster of symptoms to be compactly encompassed by a particular disorder while a layperson would not. However, in most cases, algorithms are either being deployed for expert use or to a broader population, and therefore considering the compactness for the population of interest is appropriate. Alignment differs person-by-person because different people may hold subtly or even extremely different representations of the same general concept, which affects the way they recognize and operationalize that concept. Since people are relatively tolerant of slight misalignment, this fluctuation should not have extreme consequences unless there is extreme divergence in how the concept is interpreted across the population.

\subsection{Implications}

\subsubsection{Social Media Algorithm Understanding and Folk Theories}

When algorithms meet all three ACA criteria, the choice of a mental model is simple: just the ACA concept underlying the algorithm behavior. Folk theorization requires less effort because there is a straightforward and useful mental model to deploy. However, once one or more criteria are violated, such a clean and concise mental model is no longer an option. Users are forced to make tradeoffs guided by their goals and informational contexts, resulting in a variety of folk theories which fail at effectively predicting behavior in different ways. Since most social media algorithms do not comply with the ACA criteria, users cannot satisfy all criteria at the same time, and therefore instead produce a variety of imperfect theories, mirroring the multiplicity of flawed but situationally useful theories we commonly see~\cite{eslami2016like}. Even within individual folk theories, we can explain choice of that theory in terms of particular tradeoffs between ACA criteria, satisfying one or two at the cost of another. Notably, we highlight that users are often forced to compromise on alignment (effectively, the accuracy of their mental model) because of their inability to produce non-available or non-compact theories.

In our study, users overwhelmingly relied on available and salient features to produce their theories, even at extreme cost to alignment. Our results align with earlier findings that users tend heavily toward folk theories about engagement~\cite{eslami2016like}, which are often transmitted virally through algorithmic gossip~\cite{bishop2019gossip} to reach broad audiences. Participants overwhelmingly incorporated features such as likes, replies, and views, regardless of the actual incorporation of those features into the algorithm condition, prioritizing availability over alignment. At a higher level, this phenomenon aligns with the outcomes in other research that people's expectations for behavior may override or color their perception of the actual behavior~\cite{vaccaro2018placebo} and that users incorporate and heavily weight exogenous sources (i.e., sources outside of the algorithmic system itself)~\cite{devito2018form}. At the same time, folk theories need not be particularly available to everyone, as useful theories may be learned through through communication with other users~\cite{bishop2019gossip, devito2018form}, or even decoded through coordinated effort~\cite{xiao2025influence}.

Other common folk theories similarly trade off ACA criteria depending on their origin and purpose. Some distill complex and inconsistent behavior into a straightforward and compact model like ``Algorithm will display what [platform] wants users to see''~\cite{devito2017riptwitter}, which is not particularly aligned for prediction but facilitates retroactive sensemaking and maintains compactness. 
This sort of vague theorization corresponds with the type of theorization referred to in the folk theory literature as ``causal powers'': ascribing a particular (often broad) effect to the algorithm operations. These immature theories may have local utility, and often allow users to make decisions on whether to stay or leave, but do not easily facilitate adaptation or strategic interaction with the platform~\cite{devito2021adaptive}.

Users also often hold multiple competing folk theories~\cite{eslami2016like, devito2018form}, seemingly violating compactness. 
When our participants reported non-compact mental models, they described error-patching motivations and displayed limited predictive ability. We suggest that users may enlist a multiplicity of folk theories not out of earnest belief in the validity of each, but out of an enduring uncertainty about algorithm behavior: when unable to reconcile all algorithm behavior using a single theory, they maintain a superposition of theories from which one may be retroactively selected to explain their experience. Users thereby avoid fully compromising on compactness or alignment through this strategy: they maintain a set of individually compact theories that are not aligned alone but could plausibly be so if combined in the correct manner. They can avoid having to handle the complexity of how the different components combine while also not acknowledging the lack of alignment for any particular combination or single feature. In exchange, these theory compilations are difficult to deploy for prediction due to lack of specificity.
The composite theories we encountered align with the mechanistic fragments style of theorization, where users theorize concrete details about different components of the system behavior, but not of the behavior as a whole~\cite{devito2021adaptive, keil2012running}. Mechanistic fragments are seen as often sufficient for algorithmic literacy and user decision-making on platforms~\cite{devito2021adaptive}. However, we found that users sometimes have undue confidence in incorrect fragmented theories. We believe that fragments, in contrast to a holistic understanding, make theory falsification more difficult, and therefore produce overconfidence. A \textit{correct} fragmented theory of the algorithm behavior may be appropriate to guide users' interactions with the system, but users may unknowingly maintain use of flawed fragments when they never coalesce these fragments into a holistic picture that is thoroughly tested.

\subsubsection{Designing Algorithms for ACA Compliance}

Our theory demonstrates the possibility for extremely complex algorithms to be highly predictable to users. 
In the realm of social media, suggestions like value-based feeds~\cite{jia2024embedding} are not just technically feasible and societally advantageous, but also could plausibly introduce an improvement in user understanding as compared to current feeds as well. Potential tools could allow for end-user authoring or construction of feed algorithms using conceptual building blocks (following the example of systems like Model Sketching~\cite{lam2023sketch}), but such systems should incorporate additional scaffolding to help users understand the result of combining multiple concepts, helping them reckon with compactness violations.

Our current feeds, however, remain opaque and confusing to users~\cite{eslami2016like}. Our theory provides some potential interventions towards this problem. We hypothesize that much of the confusion comes from users not understanding how the nebulous concept of ``engagement'' is operationalized. Availability is not the problem for engagement algorithms; these algorithms are well-known to exist and engagement features are displayed on the post. The issues for prediction come at the axes of compactness and alignment: while people are familiar with the general concept of engagement, they are not familiar with the particular complicated combination of features that represents ``engagement'' on a given platform, and therefore either have a scattered, piecemeal idea or a broad and nonspecific one. An effective intervention would help users develop a compact and aligned conception of what engagement means for that algorithm, perhaps by offering a concealed score or giving a more thorough accounting of how different features and outcomes are weighted, at a higher level of precision than platforms currently do. However, platforms may not care to take this step, since it facilitates users ``gaming'' the system~\cite{pierce2019paternalism}.

\subsubsection{Designing Interpretability and Explanation Approaches}

Our theory suggests that the characteristics of availability, compactness, and alignment are productive axes from which to approach user understanding methods. 

Availability is particularly fruitful, because it can be addressed both internally--in terms of the algorithm itself--and externally--in the form of communication with the user. Model architectures that are designed around concepts~\cite{koh2020conceptbottleneckmodels} are one avenue towards making the algorithm behavior more legible. However, the choice of concepts to model, as well as the features used, is itself part of the question of availability. Interpretability should therefore also concern itself with the overall design and objective of algorithms--focusing on whether the intended behavior is something that users are familiar with. However, even if the intended algorithm behavior is new, proper system design can help rectify this problem. Designers can decide on intended mental models (that are compact and aligned, as well) and help to make them more available by focusing on communicating it throughout the design~\cite{payne2007mental}. They may also opt towards straightforward user education, explaining the intentions and edge cases in a manner similar to global explanations. Broader algorithm literacy interventions may also improve the utility of these approaches, as they have been found to improve the effect of feature-based explanations which make algorithm signals more available~\cite{moon2025effects}. However, designers could also take up the expanded mandate of folk theories: including information about data collection practices and intentions of the designer as part of behavior explanations~\cite{french2017folk}. These details, while less directly representing individual model operations, are easier for users to reason about since they relate to ideas that users are already familiar with.

Concept-based model architectures also lend themselves towards improving compactness. However, if models involve many concepts that cannot be united, the improved interpretability of the internal operations will be immaterial for general understanding. A major implication of our work is how difficult users find handing even relatively few concepts, particularly in terms of understanding how they are combined or relate to each other. Even simple operations increase the difficulty for users to recognize and deploy a mental model. 

On the alignment side, the implications are more straightforward. An increase in model complexity may be justified quite frequently by the corresponding increase in accuracy. Designers should instead concern themselves with accompanying decreases in compactness, for example if the error boundary becomes less coherent~\cite{bansal2019beyondaccuracy}. Particularly fruitful are efforts to simplify the error boundary that also produce better and more generalizable models~\cite{ross2017rightrightreasonstraining}. Even when increasing accuracy towards the main construct of interest is not possible, choosing a more accurate model that represents a similar but separate concept may be preferred.  It is important for the algorithm behavior to either align with user expectations or differ from expectations in a consistent way that can become expected~\cite{hong2020human, vafa2024llmhuman, schafer2025uncertain}. Even if the models are wrong more often, people perform better when they can calibrate their trust appropriately, from the model having consistent and human-legible capability~\cite{vafa2024llmhuman, schafer2025uncertain}. Still, \textit{perfect} accuracy is not essential, especially in contexts like social media, where users are exposed to such a high volume of algorithm behavior that some errors do not preclude the development of an accurate mental model. 

Accuracy is not the only concern: algorithms that model ambiguous concepts will remain confusing when a user's particular understanding differs from the implementation. Personalization, if done well, then offers a way to make models of more subjective concepts align themselves to the expectations of users, especially because users seem to already expect it.

\section{Conclusion}

In this work, we derive ACA, a theory articulating three criteria---availability, compactness, and alignment---that together determine whether people can predict algorithmic behavior. Through an experiment, we demonstrate that users can form accurate predictive mental models for algorithms if and only if they fulfill these criteria. This theory helps explain why users occasionally find highly complex algorithms intuitive, even if their internal operations are opaque. We demonstrate the promise of a new class of algorithms, in social media feed contexts and beyond, which can meet users' needs for understanding behavior but maintain high performance towards other goals as well. 



\begin{acks}
We thank our anonymous reviewers as well as our colleagues Dora Zhao, Michelle Lam, Jonne Kamphorst, Joon Sung Park, and Jacy Reese Anthis for their thoughtful feedback on this manuscript. Our research was supported by NSF Award IIS-2403435, the Hoffman-Yee Research Grants at Stanford Institute for Human-Centered Artificial Intelligence (HAI), and the Stanford HAI-Hasso Plattner Institute Joint Research Program. Lindsay Popowski was supported by the Google PhD Research Fellowship and the Stanford Interdisciplinary Graduate Fellowship.   

\end{acks}

\bibliographystyle{ACM-Reference-Format}
\bibliography{main}
\appendix
\section{Appendix}

\subsection{ACA Measures} \label{measures}

We see opportunities for continuous measures for each of the criteria, in terms of a particular representation for the given algorithm (in most cases, we recommend using the algorithm objective, which proxies as the ideal algorithm function). For availability, one could take inspiration from lexical availability measures and ask users to generate lists of plausible algorithms or algorithm components (e.g., features) based on the exposure feed, and then construct a measure based on how frequently users described components of the algorithm. For compactness, minimum description length represents the level of cognitive complexity of the algorithm's intended behavior. And for alignment, this can be measured directly by measuring the ``accuracy'' of the algorithm against individual user annotations for the algorithm's intended behavior. Alternatively, one could ask expert coders to rank algorithms and create an ordinal measure for each criteria. Each of these approaches introduces subjectivity: the continuous measures in the selection of task, and the expert ranking through the use of subjective ratings as its basis.

\subsection{Study Procedure}\label{procedure}

\begin{figure}[htb]
  \centering
  \begin{subfigure}[t]{0.80\textwidth}
    \centering
    \includegraphics[width=\textwidth]{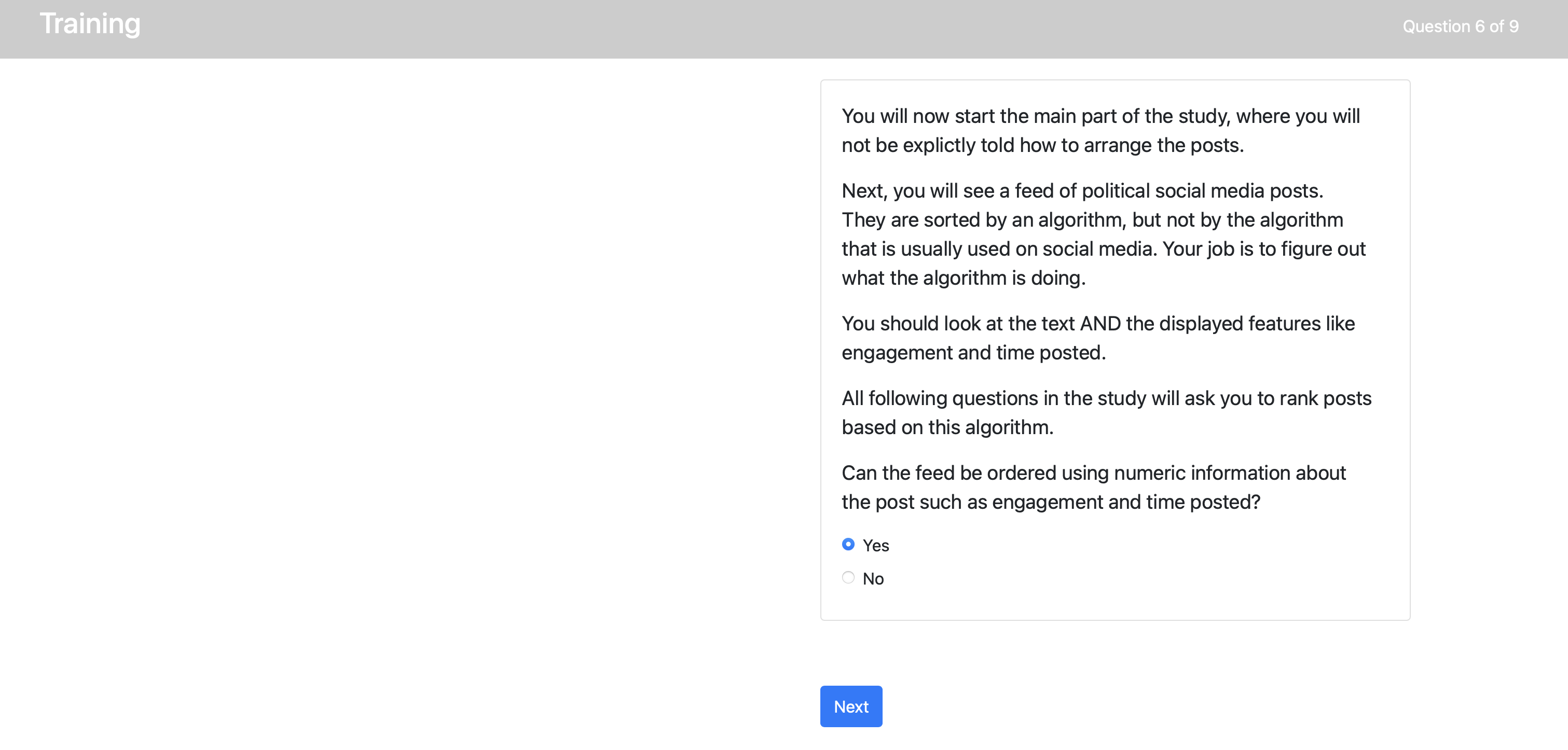}
    \Description[User instructions on the study, telling them which features to consider and testing them on comprehension.]{User instructions on the study, telling them to consider the text and displayed features to figure out what the ranking is. Later questions will use that to rank new posts. The question asks if the feed can be ordered with numeric feature counts.}
    \caption{First of four training questions to check participants' task understanding.}
    \label{subfig: ui 1}
  \end{subfigure}

\begin{subfigure}[htb]{0.80\textwidth}
    \centering
    \includegraphics[width=\textwidth]{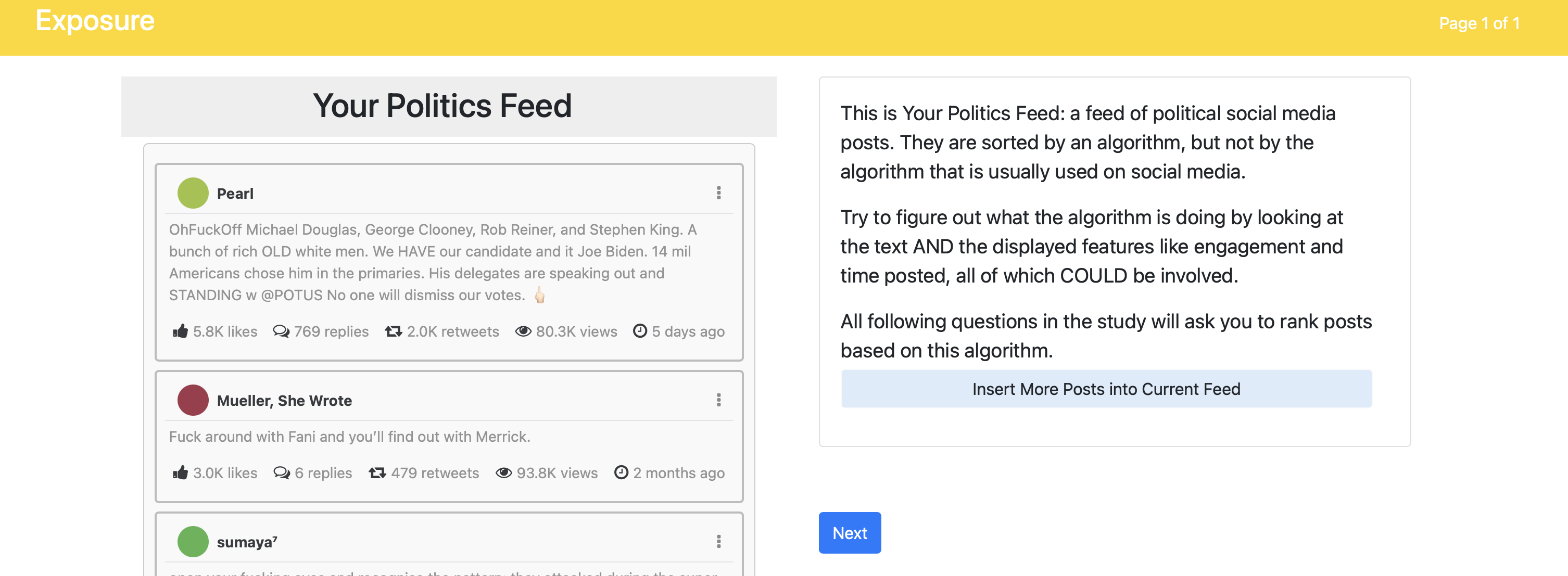}
    \Description[Example feed exposure screen with instructions to figure out the ranking.]{A feed of posts on the left side of the screen is accompanied by instructions to look at the feed and figure out what the feed is ranked by, which could include engagement, text, and time posted. A button allows users to insert more posts.}
\caption{Exposure screen with instructions for studying the feed algorithm. User can choose to add more posts. Not shown is a banner halfway down the feed reading ``Posts below this point would be ranked so low that, in a typical social media feed, it is unlikely they would ever be seen''}
    \label{subfig: ui 2}
  \end{subfigure}

\caption{Screenshots of the interface from the pre-training and exposure phases of the web experiment survey.}
  \label{fig: study ui 1}
\end{figure}

\begin{figure}[htb]
  \centering

\begin{subfigure}[htb]{0.80\textwidth}
    \centering
    \includegraphics[width=\textwidth]{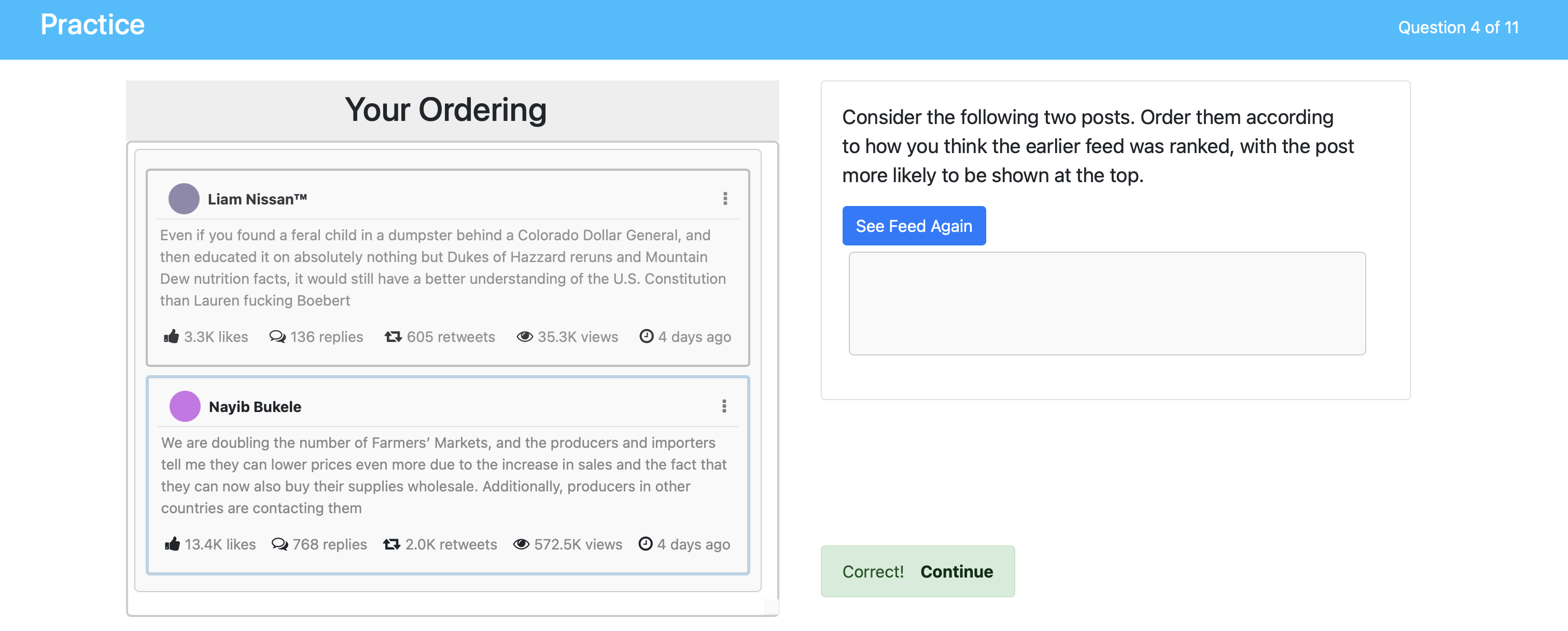}
    \Description[Example pairwise question for practice.]{Practice screen: Two posts are ranked under the heading Your Ordering. Instructions to the side tell the user to order them how the earlier feed would have. Feedback below tells the user their ranking is correct. Another button allows them to see the old feed again.}
\caption{An example practice question that requires the participant to choose which of two posts to rank higher. The feedback reports whether their ranking was correct or not.}
    \label{subfig: ui 3}
  \end{subfigure}

\begin{subfigure}[htb]{0.80\textwidth}
    \centering
    \includegraphics[width=\textwidth]{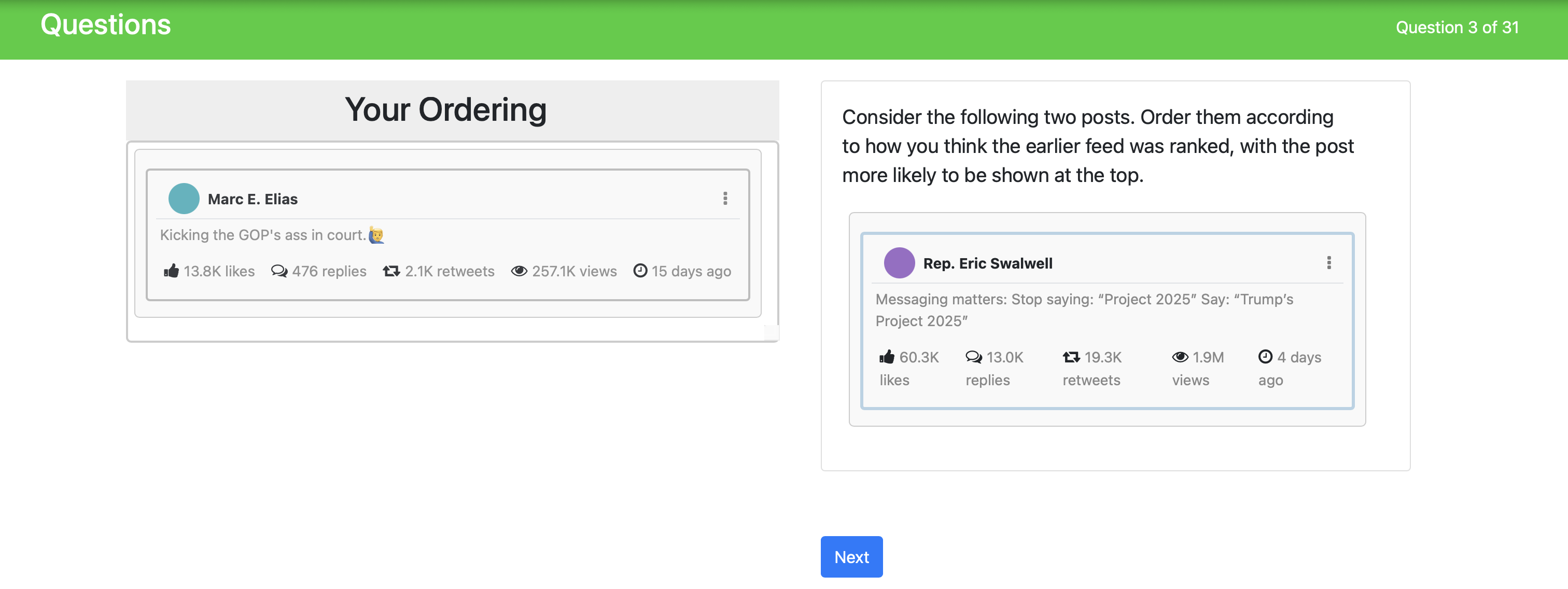}
    \Description[Example pairwise question for testing.]{Questions screen: The user is presented with two posts to order. Instructions to the side tell the user to order them how the earlier feed would have.}
\caption{An example test question that requires the participant to choose which of two posts to rank higher. No feedback is given.}
    \label{subfig: ui 4}
  \end{subfigure}

\caption{Screenshots of the interface from practice and test phases of the web experiment survey.}
  \label{fig: study ui 2}
\end{figure}

Our study has five sections: Screening, Exposure, Practice, Test, and Reflection.  We include screenshots of the UI in Figures~\ref{fig: study ui 1} and \ref{fig: study ui 2}. Participants are shown a first example task in which they are told to rank two posts by which has a larger number in the text field and are shown a second example task in which they are told to rank two posts by which is more likely to be true (e.g., one post says ``The sky is blue'' and the other says ``The sky is red''; the former should be ranked higher). They are given feedback on their correctness. If they get both questions incorrect, they are asked to leave the study. Participants are then asked a series of questions about what kinds of algorithms they may experience (exact text below).

In the exposure phase, participants are shown a feed of political posts corresponding to a particular algorithmic condition. The feed initially has 12 posts (6 highly ranked, 6 of low ranking\footnote{We divide all posts once ranked into an ordered set of eight ``buckets'' of equal size, with the highest ranked eighth of all posts in the first bucket and lowest ranked eighth in the last. When we sample ``highly'' ranked posts, they come from bucket 1 while low ranked posts come from bucket 8}), but they may click a button to repeatedly add a random selection of 8 more posts---4 high, 4 low---throughout the feed if they wish. Participants are required to stay on the exposure page for at least one minute before they are permitted to proceed. Participants are instructed to study the feed and theorize what algorithm ordered it (exact text below). 

After exposure, participants perform a training section of ten questions, where they are provided with feedback about their correctness. For each question, we choose one high and one low-ranked post to ensure there is high contrast according to the ranking algorithm between the posts. They are able to click a button to see the exact feed that they saw during the exposure phase. On the first question, they are asked to rate their confidence on their answer on a 5-point scale from ``Not at all confident'' to ``Completely confident.''

Following practice, they start the test phase of 30 questions, without any feedback. They are again asked to rate their confidence, this time on the 1st, 10th, 20th, and 30th questions. After all the prediction questions are finished, we follow up with qualitative questions about the ranking experience and participant's mental models.
Participants are required to enter responses to these questions.

For the purpose of this study, we develop a web environment that allows us to manipulate the algorithm and posts that participants are exposed to. We can mandate certain circumstances, such as the posts being sufficiently differently ranked, or making sure the current condition does not have latent correlations with particular features in the posts used for the study.

\subsubsection{Pre-check Text} \label{study_precheck}

Participants are asked the following questions and told if they are correct or not ahead of the exposure phase:

\begin{enumerate}
    \item Can the feed be ordered using numeric information about the post such as engagement and time posted?
    \item Can the feed be ordered using the text content of the post?
    \item Can the feed be ordered using the color of the profile image?
    \item Does the feed contain only political posts, or can they be about any topic?
\end{enumerate}

The correct answers are ``yes,`` ``yes,`` ``no,`` and ``only politics.''

\subsubsection{Exposure Description} \label{exposure_descript}

This is Your Politics Feed: a feed of political social media posts. They are sorted by an algorithm, but not by the algorithm that is usually used on social media. Try to figure out what the algorithm is doing by looking at the text and the displayed features, such as engagement and time posted, all of which could be involved. All following questions in the study will ask you to rank posts based on this algorithm.

\subsubsection{Task prompt}

Consider the following two posts. Order them according to how you think the earlier feed was ranked, with the post more likely to be shown at the top.

\subsection{Condition Classification} \label{condition_justification}

For each condition, we justify its classification within the ACA criteria.

\subsubsection{Likes: ACA}
Sorting by likes involves only features that are numeric and visible on each post, making it available. It uses only a single feature, making it compact. And since likes are reported exactly, there is no ambiguity in calculating the number, making it aligned.

\subsubsection{Reverse chronological: ACA}
Sorting by recency involves only features that are reported on each post, making this algorithm available. It only uses one feature, making it compact. And since the time since posting is reported on the post, there is little\footnote{If two posts were from very similar times, they could be within the rounding error of the time reported. However, since our task involves comparing very differently ranked posts, this is not an issue.} ambiguity in calculating the number, making it aligned.

\subsubsection{Severe toxicity: ACA}
Sorting by extreme toxicity involves synthesizing the textual content of the post, which humans can easily perform. Extreme toxicity stands out since it is frequently shocking or unusual, making it highly available just from reading. The public knowledge about moderating toxic and hateful content on social media also contributes to availability. Toxic or antisocial behavior is a known concept that does not need to be subdivided, making it compact. And since the toxicity classifier works quite well on political posts, the algorithm is well-aligned. Ratings of severe toxicity come from Perspective API~\cite{lees2022perspectiveapi}.

\subsubsection{Factual: ACA}
Sorting by factual-presenting content involves synthesizing the textual content of the post, which humans can easily perform. Factual-presenting posts and opinion-based posts look very different and people are primed to think about fact-checking in the political domain, making this algorithm available. People are taught to distinguish fact from opinion, so they do not have to memorize all of the individual characteristics of how facts versus opinions are expressed, making this distinction a compact concept. This algorithm is well-aligned due to its high accuracy and the strong distinction between posts that the participants are asked to compare\footnote{Our ``factual'' algorithm codes on whether posts are presented as factual content (not the actual truth of the content). Thus, issues with detecting misinformation are not relevant to alignment for this algorithm.}.  Factual ratings are obtained by prompting GPT 4o\footnote{All prompts are reported in the appendix.} and are given from 0 to 1 with 0.1 increments.

\subsubsection{Writing quality: ACA}
Sorting by writing quality involves synthesizing the textual content of the post, which humans can easily perform. Features like proper grammar and lengthy descriptions clearly distinguish quality writing without deliberate analysis, which makes this algorithm available. People are taught how to write well, making writing quality a compact concept. This algorithm is well-aligned due to its high accuracy and the strong distinction between posts that the participants are asked to compare. Quality ratings are obtained by prompting GPT 4o and are given from 0 to 1 with 0.1 increments.

\subsubsection{Likes-Views-Retweets ratio: AcA}
Both likes and views are presented clearly on the post, making the features of this algorithm available. However, the combination of likes divided by views makes it not compact, since multiple features are being combined in an unclear way. Note that while division is not a particularly complicated way to combine factors, combining factors (especially non-additively) makes the algorithm very difficult to detect. Likes divided by views does not collapse to a single recognizable concept that can be processed as a unit. Since the algorithm is exactly likes divided by views, there is no ambiguity, so it remains highly aligned. 

\subsubsection{Split quality: AcA}

This algorithm alternates between ranking by writing quality and the inverse of writing quality depending on if the post has more than the median like count. All of the features used are available and aligned, as we argue earlier, and their combination is exact and deterministic, so it maintains alignment. However, the features are combined in a way that is overly complex, failing compactness.

\subsubsection{Split engagement: AcA}

This algorithm alternates between ranking by likes minus twenty times retweets and twenty times retweets minus likes depending on if the post is above median toxicity. All of the features used are available and aligned, as we argue earlier, and their combination is exact and deterministic, so it maintains alignment. However, the features are combined in a way that is overly complex, failing compactness.

\subsubsection{Engagement-Factual-Toxicity combination: AcA}

This algorithm combines the views, retweets, and likes with toxicity and factual presentation of a post in a highly complicated way. All of those individual features we have argued to be available and aligned, and nothing about the combination violates alignment (e.g., by adding noise or randomness). However, the combination of these features is complicated enough to violate compactness.

\subsubsection{First-person pronouns: aCA}

The count of first-person pronouns is not easily recognizable from looking at or even reading a post. This algorithm is therefore not available. The idea of first-person pronouns is a taught concept, making it compact. And the algorithm is calculated only by counting the number of these words, preventing any ambiguity since posts with different numbers are being compared. First-person pronouns is therefore aligned.

\subsubsection{Odd digits: aCA}

The proportion of odd digits in the engagement features does not stand out visually, nor is it something that readily comes to mind in the social media algorithm space. However, the proportion of odd digits can be thought of as a single concept, making it compact. Since it is calculated exactly using features displayed, it is aligned.

\subsubsection{Commas: aCA}

The proportion of commas relative to overall punctuation does not stand out visually, nor is it something that readily comes to mind in the social media algorithm space. However, the proportion of commas is a single concept, so it is compact. Since it is calculated straightforwardly using the text as shown, it is aligned.

\subsubsection{Time-related: aCA}

Sorting by how time-related a post is relies only on the post text, which humans are good at processing. However, ranking by the relevancy of a post to the concept of time is in no way expected for the social media context, nor is it made readibly visible, making it unavailable. It is a singular concept, making it compact. It is also highly accurate and aligned with a human's concept of time, making it aligned in our context of comparing posts that vary highly on the axis of time-relation.

\subsubsection{Caring: ACa}

Sorting by how caring a post is relies only on the post text, which humans are good at processing. A very caring post sounds very different from an uncaring one in the political domain, and people are primed to think about kindness or positivity versus unkindness and negativity in the social media space due to news and research, making it available. Caring as a concept is well known to people, making it compact. However, this algorithm does not perform accurately enough among the political tweets we use, making it unaligned. This algorithm is implemented using a BERT-based architecture that can report the presence of different values~\cite{kiesel2022identifyingvalues}.

\subsubsection{Achievement: ACa}

Sorting by how achievement-focused a post is relies only on the post text, which humans are good at processing. Discussions of how people are rewarded on social media for sharing achievements and successes on are frequent, which we believe makes achievement an available concept in this space. Achievement is a singular concept that most people already hold, making it compact. However, this algorithm does not perform accurately enough among the political tweets we use, making it unaligned. This algorithm is implemented using the same BERT-based architecture as the Caring algorithm.

\subsubsection{Noisy likes: ACa}

We justify earlier why an algorithm sorting by most likes would be available and compact. However, in this implementation, we add noise, by randomly switching a portion of the posts' rankings with those of other posts. This step of adding noise to the algorithm makes it unaligned.

\subsubsection{Noisy factual: ACa}

We justify earlier why an algorithm sorting by how factually a post is presented would be available and compact. However, in this implementation, we add noise, by randomly switching a portion of the posts' rankings with other posts. This step of adding noise to the algorithm makes it unaligned.

\subsubsection{Grammatical features: acA}
This algorithm uses different grammatical features (10 times the ratio of second person to first person pronouns, plus the average word length, plus the number of punctuation marks) which are all not available upon looking at the post. By combining multiple different features without a clear reason for the association, it is not compact. However, when applying this algorithm, there is no ambiguity or errors, making it aligned.

\subsubsection{Split grammar: acA}

This algorithm alternates between different sets of features relating to grammar. If the post has more than median likes, it is sorted by the ratio of first person pronouns to all pronouns, otherwise, the ratio of commas to overall punctuation. We have previously classed many of these features (all but likes) as unavailable, and their complex combination renders them non-compact, but this algorithm remains aligned since the calculation maintains exactness and could be executed without introducing error.

\subsubsection{Conservative values combination: Aca}
This algorithm, which combines ratings for several concepts (tradition, achievement, personal security, and conformity to rules), uses values that are familiar to people and salient in the political context, making them available. However, by combining so many separate concepts, this algorithm becomes non-compact. Due to the lower than necessary performance of each individual classifier for the different values, this algorithm is also unaligned. The value concept rating is performed by the same BERT-based model as in the Caring condition.

\subsubsection{Noisy likes-retweets-views ratio: Aca}

This algorithm takes an algorithm that was already available but not compact, and then additionally violates alignment through adding random noise to the ranking.

\subsubsection{Noisy first-person pronouns: aCa}

This algorithm takes an algorithm that was already compact but not available, and then additionally violates alignment through adding random noise to the ranking.

\subsubsection{Wintertime: aCa}

This algorithm poorly executes on a unavailable concept to create an algorithm that is compact but not available or aligned (because in the inventory of the posts we have, winter is rarely mentioned or relevant). Relating to the concept of winter is not visually apparent nor expected in the context of politics or social media, violating availability. It is also specified unclearly in the prompt (this algorithm is implemented with GPT 4o), making it fail alignment, especially in the political context. 

\subsubsection{Noisy grammatical features: aca}

This algorithm takes an existing algorithm that we deemed to violate availability and compactness, and then additionally violates alignment by adding noise.

\subsubsection{Caring-punctuation: aca}

This algorithm combines something that violates alignment (Caring) with something that violates availability (ratio of commas to overall punctuation). And through the unusual and non-cohesive combination of these two features, the algorithm also violates compactness.

\subsection{GPT Prompts}

\subsubsection{Political:}
\texttt{Political content on Twitter is varied and can be about officials and activists, social issues, or news and current events. Looking at the following tweet, would you categorize it as POLITICAL or NOT POLITICAL content? Answer 1 if it is POLITICAL, 0 otherwise, with no explanations.}

\subsubsection{Quality:} 
\texttt{Posts on twitter may vary wildly in the quality of their argument.
Quality writing contains well-reasoned thoughts that are clear without excessive length or use of jargon, while lower-quality writing may have incomplete ideas, lack substance, context, reasoning, or factual content, or be overly verbose.
Rank the following tweet from 0 (very low quality and effort) to 1 (high quality writing) with one decimal point of precision.
Give output of the json form \{"quality": QUALITY\} with no explanations.}

\subsubsection{Factual:} 
\texttt{Posts on twitter may contain both facts and opinions. Factual content is unbiased reporting of actual events, while opinions involve interpretation of events from a particular viewpoint. 
Rank the following tweet from 0 (no factual content ) to 1 (entirely factual content) with one decimal point of precision. Give output of the json form \{"factual": FACTUAL\} with no explanations.}

\subsection{Earlier Pre-registration and Study}\label{previous}

We previously pre-registered and ran two studies\footnote{\url{https://osf.io/ukmc8} and \url{https://osf.io/n9kgd}.} using a very similar study design and a slightly different set of conditions. 

\subsubsection{Original Study Design}

In our previous study, we used three conditions not present here, Likes-views ratio (AcA, likes divided by views), random (aCa), and hashes (aca, a hash function applied to the post text). The remaining algorithm conditions were: Likes, Chronological, Severe toxicity, Factual, Writing quality, Caring, First-person pronouns, Grammatical features, and Conservative values (all present in this iteration of the study).

All other procedures were the same as our current one, except participants were required to look at the feed exposure for two minutes (rather than one, in the current iteration). Our hypothesis remained the same: that conditions classified as available, compact, and aligned would result in higher prediction accuracy among participants. Our statistical analysis was very similar (see pre-registration document). We performed slightly different qualitative analysis of the participant mental models.

\subsubsection{Original Study Results and Follow-up}

Our study results mostly aligned with our hypothesis: participants predicted with higher accuracy in the ACA conditions versus the non-ACA conditions, and the the three-way interaction effect (A*C*A) significant, corresponding to better odds of correct prediction. However, the availability-alignment effect was also positive and significant, indicating that the three-way interaction was not the only important one. This corresponded with the Likes-reviews ratio condition being predicted at an accuracy significantly above random guessing. Since this result challenged our hypothesis, we investigated deeper by analyzing the mental models, and realized that many participants used mental models unrelated to the ratio of likes to views while having high prediction accuracy. We simulated potential prediction performance on this condition when using the alternate mental models that participants had described, and found use of alternate mental models could account for the higher-than-expected prediction performance that we saw.

In order to better understand why users overperformed in the Likes-views ratio condition, we pre-registered and ran a second study designed to investigate the use of correlated algorithms in a mental model. In this study, we showed that we could lower participant prediction accuracy by picking collections of posts that did not have latent correlations with the algorithms our participants had described using as mental models.

\subsubsection{Considerations in Study Re-design}
We ultimately chose to rerun the study with twice as many conditions to demonstrate the possibility of more than one algorithm per ACA category, and to switch out three conditions for which classification was less straightforward.

We opted to remove the random condition due to its trivially compact nature (if randomness is a concept, what utility does that concept have?) and due to its similarity in implementation to hashes (which has behavior indistinguishable from random for a particular person).
We removed likes-views ratio as a condition due to its higher-than-average correlation with other available algorithms and because a two-feature algorithm existed in a bit of a compactness gray-area.

\end{document}